\documentclass{iopart}
\usepackage{graphicx}  
\usepackage{latexsym}
\usepackage{amssymb}

\jl{6}        
\eqnobysec    

\def\beq{\begin{equation}}
\def\eeq{\end{equation}}
\def\rmd{{\rm d}}

\def\rightcontract{\mathop{\hbox{\vrule width0.5pt height6pt%
  \vrule height0.5pt width6pt}}}

\def\leftcontractp#1{\mathop{\rlap{\hbox to 8pt{\hss$^{#1}$\hss}}%
  \hbox{\vrule height0.5pt width8pt \vrule width0.5pt  height6pt}}}
 
\def\rightcontractp#1{\mathop{%
  \hbox{\vrule width0.5pt height6pt \vrule height0.5pt width8pt}%
    \llap{\hbox to 8pt{\hss$^{#1}$\hss}}}}
 
\def\hook{\,\mathop{\breve{\,}}\nolimits\,}
\def\lefthook{\hook\kern-1.5pt}

\begin{document}

\title[Accelerated orbits in black hole fields: the static case]
{Accelerated orbits in black hole fields: the static case}

\author{
Donato Bini$^* {}^\S{}^\dag$,
Fernando de Felice$^\P {}^\diamond$ and
Andrea Geralico${}^\S{}^\ddag$
}

\address{${}^*$\
Istituto per le Applicazioni del Calcolo ``M. Picone,'' CNR, 
I--00185 Rome, Italy}

\address{
  ${}^\S$\
  ICRA,
  University of Rome ``La Sapienza,'' I--00185 Rome, Italy
}

\address{
  ${}^\dag$\
  INFN, Sezione di Firenze, I--00185
  Sesto Fiorentino (FI), Italy
}

\address{
${}^\P$\
University of Padova, I-35131, Padova, Italy}

\address{
${}^\diamond$\
INFN,  Sezione di Padova,   I-35131, Padova, Italy}

\address{
  $^\ddag$
  Physics Department,
  University of Rome ``La Sapienza,'' I--00185 Rome, Italy}

\begin{abstract}
We study  non-geodesic orbits of test particles endowed with a structure, assuming the  Schwarzschild spacetime as  background. 
We develop a formalism which allows one to recognize the geometrical characterization of those orbits in terms of their Frenet-Serret parameters and apply it to explicit cases as those of spatially circular orbits which witness the equilibrium under conflicting types of interactions. 
In our general analysis we solve  the equations 
of motion offering a detailed picture of the dynamics having in mind a check with a possible astronomical set up. We focus on certain ambiguities which plague the interpretation of the measurements   preventing one from identifying the particular structure carried by the particle. 
\end{abstract}

\pacno{04.20.Cv}

\section{Introduction}

Deviations from geodesic motion by a test particle in a given gravitational field are generally  associated with  particle's additional properties, like electric charge or spin, but also with external interacting fields (even test ones) added to the  background. 
Consistency with the test hypothesis of the particle and the superposed field requires small deviations from the geodesic behaviour, otherwise  back-reaction should be taken into account with awkward implications.
Indeed, assuming the consistency condition to be satisfied, a detailed study of such deviations can be easily done.

The motion of charged particles as well as particles with magnetic moment in the field of a black hole when a magnetic field is also present has been analyzed in several papers \cite{prasanna1,prasanna2,prasanna3,prasanna4,prasanna5,fdf_preti,fdf_preti2}. 
Contrary to the previous studies our approach unables us to solve the equations of motion analytically, so identifying the equilibrium solutions and the stability conditions. These properties are of clear relevance in the astrophysical context. 
Moreover, our method can be generalized to find the behaviour of bodies endowed with whatever physical structure and even with more than one being present in the same object: this is the case of bodies with both spin and magnetic moment. The results of our analysis can be directly confronted with astrophysical observations. 

In this paper, we provide an intrinsic characterization of general equatorial orbits in the Schwarzschild spacetime by studying their Frenet-Serret properties, i.e. curvature and torsions, filling a gap existing in the literature where  only  circular orbits are considered in detail.
Here, we make explicit examples of forces which are responsible of these orbits, by considering --within  a unified scenario-- charged particles and  magnetic dipoles moving in the given metric with a magnetic field added, spinning test bodies gravitationally interacting with the background and particles deviating from the geodesic behaviour by scattering of electromagnetic radiation. 

The present work can also be useful for the following reason.
Nowadays progress in the instrumental sensitivity and data analysis has allowed a careful reconstruction of the orbits of a large number of astrophysical objects and hence models can be considered to fit the data. This is the case  of pulsars orbiting  around Sgr A$^*$, the super-massive black hole at the center of our galaxy.
For instance, it is largely believed that to measure general relativistic effects with current instruments, one needs to discover pulsars with orbital periods less than about 10 years and orbiting  at a distance of about  $0.01$ pc from Sgr A$^*$. Even if searches are currently underway, at least five pulsars are now known to lie within $100$ pc from Sgr A$^*$ \cite{psr1}.
Reconstruction of the orbits directly from observations is not so far from becoming possible.
Hence, if a particle deviates from geodesic motion  within a properly chosen background spacetime and the deviations are measured, one may use such measurements to put limits on the particle structure or to the nature and  intensity of external fields. 
We make this point explicit in the simple context of a Schwarzschild black hole.

Finally, another outcome of this work is the discussion of analogies in the motion of differently structured particles. Under certain conditions it may happen that either an electrically charged particle behaves in an external electromagnetic field  exactly like a neutral spinning one in the given background or a spinning particle endowed with a magnetic dipole moment in an external magnetic field becomes indistinguishable from a non-spinning and neutral particle with suitable values of the above properties.
 
Hereafter latin indices run from $1$ to $3$ whereas greek indices run from $0$ to $3$ and geometrical units are assumed. The metric signature is chosen 
as $+2$.

\section{The intrinsic properties of equatorial orbits}

Schwarzschild metric, expressed in standard coordinates $(t,r,\theta,\phi)$, is given by
\beq
\label{Schwmetric}
\rmd  s^2 = -N^2\rmd t^2 + N^{-2} \rmd r^2 
+ r^2 (\rmd \theta^2 +\sin^2 \theta\, \rmd \phi^2)\,,
\eeq
where  
$N=\left(1-{2M}/r\right)^{1/2}$ is the lapse function. It is understood that the following discussion only holds outside the horizon, namely for $r>2M$.
The unit volume $4$-form which assures the orientation of the spacetime is denoted by $\eta_{\alpha\beta\gamma\delta}$ and related to the Levi-Civita alternating symbol $\epsilon_{\alpha\beta\gamma\delta}$ ($\epsilon_{0123}=1$) by the relation
\beq
\eta_{\alpha\beta\gamma\delta}=\sqrt{-g} \epsilon_{\alpha\beta\gamma\delta}\,,
\eeq
where $g$ is the determinant of the metric.

Introduce the standard orthonormal frame adapted to the static observers (in the case of metric (\ref{Schwmetric}) they coincide with ZAMOs, Zero Angular Momentum Observers), namely
\beq 
\label{frame}
\fl\qquad
e_{\hat t}\equiv n=N^{-1}\partial_t\,, \quad
e_{\hat r}=N\partial_r\,, \quad
e_{\hat \theta}={r}^{-1}\partial_\theta\,, \quad
e_{\hat \phi}=({r\sin \theta})^{-1}\partial_\phi\,.
\eeq
Let $U$ be the $4$-velocity  of a test particle,
\beq
\label{U4vel}
\fl\quad
U=\gamma (n +\nu^{\hat r}e_{\hat r}+\nu^{\hat \theta}e_{\hat \theta}+\nu^{\hat \phi}e_{\hat \phi})\,,\qquad
\gamma=1/\sqrt{1-\nu^2}\,, \qquad
\nu^2=\delta_{\hat a \hat b}\nu^{\hat a}\nu^{\hat b}\,,
\eeq
where $\nu^{\hat a}$ ($a=r,\theta, \phi$) are the components of the spatial velocity relative to $n$ and $\gamma$ is the Lorentz factor.
Its $4$-acceleration frame components are given by
\begin{eqnarray}
\label{fouracc}
a(U)^{\hat t} 
&=& \frac{d\gamma}{d\tau}+\frac{\gamma^2N}{r}\nu_K^2\nu^{\hat r}\,,\nonumber \\
a(U)^{\hat r}
&=& \frac{d (\gamma \nu^{\hat r})}{d\tau}-\frac{\gamma^2 N}{r}(\nu^{\hat \theta}{}^2+\nu^{\hat \phi}{}^2-\nu_K^2)\,,\nonumber \\
a(U)^{\hat \theta}&=& \frac{d (\gamma \nu^{\hat \theta})}{d\tau}+\frac{\gamma^2}{r\sin\theta}[N\sin \theta \nu^{\hat r}\nu^{\hat \theta}-\cos \theta \nu^{\hat \phi}{}^2]\,,\nonumber \\
a(U)^{\hat \phi}&=& \frac{d (\gamma \nu^{\hat \phi})}{d\tau}+\frac{\gamma^2}{r\sin\theta}[N\sin \theta \nu^{\hat r}+\cos \theta \nu^{\hat \theta}]\nu^{\hat \phi}\,,
\end{eqnarray}
where 
\beq
\nu_K^2=\frac{1-N^2}{2N^2}=\frac{M}{r-2M}
\eeq
is the speed of  a Keplerian spatially circular geodesic at radius $r$ with associated Lorentz factor $\gamma_K=1/\sqrt{1-\nu_K^2}$.
From the orthogonality condition $U\cdot a(U)=0$ the following relation  holds
\beq
\label{formula_acc}
a(U)^{\hat t}=\nu_{\hat r} a(U)^{\hat r}+\nu_{\hat \theta} a(U)^{\hat \theta}+\nu_{\hat \phi} a(U)^{\hat \phi}\,.
\eeq

We shall now study  the special case of a motion confined to the equatorial plane ($\theta=\pi/2$), namely with $\nu^{\hat \theta}=0$;  the above relations then simplify as follows
\begin{eqnarray}
\label{fouracceq}
a(U)^{\hat t}
&=& \frac{d\gamma}{d\tau}+\frac{\gamma^2N}{r}\nu_K^2\nu^{\hat r}\,,\nonumber \\
a(U)^{\hat r}
&=& \frac{d (\gamma \nu^{\hat r})}{d\tau}-\frac{\gamma^2 N}{r}(\nu^{\hat \phi}{}^2-\nu_K^2)\,,\nonumber \\
a(U)^{\hat \theta}&=&0\,,\nonumber \\
a(U)^{\hat \phi}&=& \frac{d (\gamma \nu^{\hat \phi})}{d\tau}+\frac{\gamma^2 N }{r} \nu^{\hat r}\nu^{\hat \phi}\,,
\end{eqnarray}
with also, from Eq. (\ref{U4vel}),
\begin{eqnarray}
\frac{\rmd r}{\rmd \tau}&=& \gamma N \nu^{\hat r}\,,\qquad
\frac{\rmd \phi}{\rmd \tau}=\frac{\gamma}{r} \nu^{\hat \phi}\,.
\end{eqnarray}
The general equatorial motion is thus fully described by the two equations
\begin{eqnarray}
\label{eq_nus}
\frac{d \nu^{\hat r}}{d\tau}
&=& \frac{1}{\gamma}\left(\frac{a(U)_{\hat r}}{\gamma_r^2}- a(U)_{\hat \phi}\nu^{\hat r}\nu^{\hat \phi}\right)+
\frac{\gamma N}{r}[\nu^{\hat \phi}{}^2 -\nu_K^2 (1-\nu^{\hat r}{}^2)]\,,\nonumber \\
\frac{d  \nu^{\hat \phi}}{d\tau}
&=& 
\frac{1}{\gamma}\left(\frac{a(U)_{\hat \phi}}{\gamma_\phi^2}- a(U)_{\hat r}\nu^{\hat r}\nu^{\hat \phi}\right)
-\frac{\gamma N}{\gamma_K^2 r}\nu^{\hat r}\nu^{\hat \phi}\,,
\end{eqnarray}
where $\gamma_r=1/\sqrt{1-\nu^{\hat r}{}^2}$ and $\gamma_\phi=1/\sqrt{1-\nu^{\hat \phi}{}^2}$.
Here all the components of $a(U)$ have been re-expressed in terms of  $a(U)_{\hat r}$ and $a(U)_{\hat \phi}$, by using Eqs. (\ref{formula_acc})--(\ref{fouracceq}), so that
\beq
\label{a_di_U}
a(U)=[\nu_{\hat r} a(U)^{\hat r}+\nu_{\hat \phi} a(U)^{\hat \phi}]n+a(U)^{\hat r}e_{\hat r}+ a(U)^{\hat \phi}e_{\hat \phi}\,,
\eeq
and
\begin{eqnarray}\fl\quad
\label{kappa_eq}
||a(U)||\equiv \kappa&=&\sqrt{-[a(U)_{\hat r}\nu^{\hat r}+a(U)_{\hat \phi}\nu^{\hat \phi}]^2+a(U)_{\hat r}{}^2+a(U)_{\hat \phi}{}^2}\nonumber\\
\fl\quad
&=& \left[\frac{a(U)_{\hat r}^2}{\gamma_r^2}+\frac{a(U)_{\hat \phi}^2}{\gamma_\phi^2}-2\nu^{\hat r}\nu^{\hat \phi}a(U)_{\hat r}a(U)_{\hat \phi}\right]^{1/2}\nonumber\\
\fl\quad
&=& \gamma^2\left\{\frac{\dot\nu^{\hat r}{}^2}{\gamma_\phi^2}+\frac{\dot\nu^{\hat \phi}{}^2}{\gamma_r^2}+2\nu^{\hat r}\nu^{\hat \phi}\dot\nu^{\hat r}\dot\nu^{\hat \phi}-\frac{2}{\gamma}\frac{N}{r}[(\nu^{\hat \phi}{}^2-\nu_K^2)\dot\nu^{\hat r}-\nu^{\hat r}\nu^{\hat \phi}\dot\nu^{\hat \phi}]\right.\nonumber\\
\fl\quad
&&\left.
+\frac{N^2}{r^2}[(\nu^{\hat \phi}{}^2-\nu_K^4)\nu^{\hat r}{}^2+(\nu^{\hat \phi}{}^2-\nu_K^2)^2]
\right\}^{1/2} 
\,,
\end{eqnarray}
where a dot stands for differentiation with respect to the proper time.

In the equatorial plane, the condition for a spatially circular motion, namely $r=r_0=$ const., $\nu^{\hat r}=0$ and ${d \nu^{\hat r}}/{d\tau}=0$,  simplifies the equations of motion which become
\beq
\label{f_di_r}
\frac{a(U)_{\hat r}}{\gamma}+\gamma\frac{N}{r}(\nu^{\hat \phi}{}^2 -\nu_K^2)=0\,,\qquad
\frac{d  \nu^{\hat \phi}}{d\tau}=\frac{a(U)_{\hat \phi}}{\gamma^3}    \,.
\eeq
If in addition $a(U)_{\hat \phi}=0$ then $\nu^{\hat \phi}=$ const. and  $a(U)_{\hat r}=$ const. as can be seen from the first of Eq. (\ref{f_di_r}), namely
\beq
\label{sol_circ}
a(U)_{\hat r}=-\gamma^2\frac{N}{r}(\nu^{\hat \phi}{}^2 -\nu_K^2)\,.
\eeq

Going back to the general case, the vectors  of the ZAMO frame have   transport laws along $U$ given by
\begin{eqnarray}
\label{vec_fond}
\fl\quad
\nabla_Un=\gamma \frac{N}{r}\nu_K^2e_{\hat r}\,,\quad
\nabla_Ue_{\hat r}=\gamma \frac{N}{r}\left(\nu_K^2n+\nu^{\hat\phi}e_{\hat \phi}\right)\,,\quad
\nabla_Ue_{\hat \phi}=-\gamma \frac{N}{r}\nu^{\hat\phi}e_{\hat r}\,.
\end{eqnarray}
In order to deduce the intrinsic properties of the orbits under consideration, we shall
set up a Frenet-Serret frame $\{E_\alpha\}$ adapted to the orbit, i.e. with $E_0=U$ and satisfying the standard relations
\beq
\label{FSeqs}
\fl\qquad
\begin{array}{llllll}
\nabla_U E_0&=&\kappa E_1\,, & \qquad
\nabla_U E_1&=&\kappa E_0+\tau_1 E_2\,, \\
\nabla_UE_2&=&-\tau_1E_1+\tau_2E_3\,, & \qquad
\nabla_UE_3&=&-\tau_2E_2\,, 
\end{array}
\eeq
where $\tau_1$ and $\tau_2$ are the first and the second torsion respectively.
Recalling Eqs. (\ref{a_di_U})--(\ref{kappa_eq}) we have
\beq\fl\qquad
\label{E1def}
E_1 \equiv \frac{a(U)}{||a(U)||}
=\frac{1}{\kappa}\left[(a(U)_{\hat r}\nu^{\hat r}+a(U)_{\hat \phi}\nu^{\hat \phi})\,n+a(U)^{\hat r}e_{\hat r}+a(U)^{\hat \phi}e_{\hat \phi}\right]\,.
\eeq
Following the Frenet-Serret procedure we find
\begin{eqnarray}\fl\qquad
\label{E23def}
E_2&=&\frac{\gamma}{\kappa}\left[
(\nu^{\hat \phi}a(U)_{\hat r}-\nu^{\hat r}a(U)_{\hat \phi})n+
\left(\nu^{\hat \phi}\nu^{\hat r}a(U)_{\hat r}-\frac{a(U)_{\hat \phi}}{\gamma_\phi^2}\right)e_{\hat r}\right.\nonumber\\
\fl\qquad
&&\left.-\left(\nu^{\hat \phi}\nu^{\hat r}a(U)_{\hat \phi}-\frac{a(U)_{\hat r}}{\gamma_r^2}\right)e_{\hat \phi}
\right]\,,\nonumber\\
\fl\qquad
E_3&=&-e_{\hat \theta}\,.
\end{eqnarray}
The magnitude of the 4-acceleration is defined by Eq. (\ref{kappa_eq}); the first torsion is given by
\begin{eqnarray}
\label{tau1_def}\fl
\kappa^2\tau_1
&=&\frac{N}{r}\left[\nu^{\hat \phi}a(U)_{\hat r}^2-\nu_K^2\nu^{\hat r}a(U)_{\hat r}a(U)_{\hat \phi}+\frac{1}{\gamma_K^2}\nu^{\hat \phi}a(U)_{\hat \phi}^2\right]\nonumber\\
\fl
&&+\frac{1}{\gamma}\left(a(U)_{\hat r}\frac{da(U)_{\hat \phi}}{d\tau}-a(U)_{\hat \phi}\frac{da(U)_{\hat r}}{d\tau}\right)+\kappa^2 (\nu^{\hat r}a(U)_{\hat \phi}-\nu^{\hat \phi}a(U)_{\hat r})\nonumber\\
\fl
&=&\frac{N}{r}\frac{\gamma^2}{\gamma_K^2}\kappa^2\nu^{\hat \phi}-\gamma^3\left\{
\ddot\nu^{\hat r}\left[\dot\nu^{\hat \phi}+\frac{N}{r}\frac{\gamma}{\gamma_K^2}\nu^{\hat r}\nu^{\hat \phi}\right]
-\ddot\nu^{\hat \phi}\left[\dot\nu^{\hat r}-\frac{N}{r}\gamma\left(\nu^{\hat \phi}{}^2-\frac{\nu_K^2}{\gamma_r^2}\right)\right]
\right\}\nonumber\\
\fl
&&+ \frac{N}{r}\gamma^4\left[\nu^{\hat \phi}\left(\frac{\dot\nu^{\hat r}{}^2}{\gamma_K^2}+2\dot\nu^{\hat \phi}{}^2\right)+(1+\nu_K^2)\nu^{\hat r}\dot\nu^{\hat r}\dot\nu^{\hat \phi}\right]
+\frac{N^3}{r^3}\gamma^4\nu^{\hat \phi}\nu^{\hat r}{}^2\nu_K^2(1+2\nu_K^2)\nonumber\\
\fl
&&-\frac{N^2}{r^2}\gamma^5\left\{
\nu^{\hat \phi}\dot\nu^{\hat r}\left[\frac1{\gamma_K^2}(\nu^{\hat \phi}{}^2-\nu_K^2)+\nu^{\hat r}{}^2(1-3\nu_K^2-\nu_K^4)\right]\right.\nonumber\\
\fl
&&\left.+\nu^{\hat r}\dot\nu^{\hat \phi}\nu_K^2\left(\nu^{\hat \phi}{}^2-\frac{3+\nu_K^2}{\gamma_r^2}\right)
\right\}\,.
\end{eqnarray}
The second torsion $\tau_2$ vanishes identically due to the fact that the orbit lies in the equatorial plane and the Schwarzschild metric is reflection symmetric about that plane.

Note that if $\nu^{\hat r}=0$, $\nu^{\hat \phi}=$ const. (spatially circular orbits)  we have \cite{Iyer:1993qa}
\beq
\label{tau1_circ}
\tau_1=\frac{N}{r}\frac{\gamma^2}{\gamma_K^2}\nu^{\hat \phi}\,,\qquad \kappa 
=\frac{N}{r}\gamma^2 |\nu^{\hat \phi}{}^2-\nu_K^2|\,.
\eeq
One can also consider a signed magnitude of $\kappa$, i.e.
\beq
\kappa_{\rm (sm)}\equiv -\frac{N}{r}\gamma^2 (\nu^{\hat \phi}{}^2-\nu_K^2)\,,
\eeq
which, once differentiated with respect to $\nu^{\hat \phi}$, has the following  expression in terms of the first torsion
\beq
\frac{d \kappa_{\rm (sm)}}{d\nu}=-2\frac{N}{r} \frac{\gamma^4 }{\gamma_K^2}\nu^{\hat \phi}=-2\gamma^2 \tau_1\,.
\eeq

\section{Special cases of acceleration}

We shall now examine some special cases.

\noindent
{\bf  \underline{Case 1}: Spatially radial acceleration, $a(U)_{\hat t}\not =0$, $a(U)_{\hat \phi}=0$}

In the case $a(U)_{\hat \phi}=0$, i.e.
\beq
a(U)=a(U)_{\hat r}\left(\nu^{\hat r}\,n+e_{\hat r}\right)\,,\qquad
\kappa=\frac{|a(U)_{\hat r}|}{\gamma_r}\,,
\eeq
the Frenet-Serret frame (\ref{E1def})--(\ref{E23def}) reduce to
\begin{eqnarray}
E_1 &=&\epsilon_r\gamma_r\left(\nu^{\hat r}\,n+e_{\hat r}\right)\,,\qquad 
\epsilon_r={\rm sgn}[a(U)_{\hat r}]\,,\nonumber\\
E_2&=&\epsilon_r\gamma\gamma_r\left[
\nu^{\hat \phi}(n+\nu^{\hat r}e_{\hat r})+\frac{1}{\gamma_r^2}e_{\hat \phi}
\right]\,,\nonumber\\
E_3&=&-e_{\hat \theta}\,,
\end{eqnarray}
and the first torsion (\ref{tau1_def}) becomes
\beq
\tau_1=\nu^{\hat \phi}\left(\gamma_r^2\frac{N}{r}- a(U)_{\hat r}\right)\,.
\eeq

This particular situation also includes the case of spatially circular orbits where in addition $\nu_{\hat r}=0$ and $\gamma_r=1$, so that 
\beq\fl\qquad
a(U)=a(U)_{\hat r}e_{\hat r}\,,\qquad 
\kappa=|a(U)_{\hat r}|\,,\qquad
\tau_1=\nu^{\hat \phi}\left(\frac{N}{r}- a(U)_{\hat r}\right)\,,
\eeq
and 
\beq
E_1=\epsilon_r e_{\hat r}\,,\qquad
E_2=\epsilon_r\gamma (\nu^{\hat \phi} n +e_{\hat \phi})\,,\qquad
E_3=-e_{\hat \theta}\,.
\eeq

\noindent
{\bf \underline{Case 2}:  Spatially tangential acceleration, $a(U)_{\hat t}\not=0$, $a(U)_{\hat r}=0$}

In the special case $a(U)_{\hat r}=0$, i.e.
\beq
a(U)=a(U)_{\hat \phi}\left(\nu^{\hat \phi}\,n+e_{\hat \phi}\right)\,,\qquad
\kappa =\frac{|a(U)_{\hat \phi}|}{\gamma_\phi}\,,
\eeq
the Frenet-Serret relations reduce to
\begin{eqnarray}
E_1 &=&\epsilon_\phi \gamma_\phi\left(\nu^{\hat \phi}\,n+e_{\hat \phi}\right)\,,\qquad 
\epsilon_\phi={\rm sgn}[a(U)_{\hat \phi}]\,,\nonumber\\
E_2&=&-\epsilon_\phi \gamma\gamma_\phi\left[
\nu^{\hat r}(n+\nu^{\hat \phi}e_{\hat \phi})+\frac{1}{\gamma_\phi^2}e_{\hat r}
\right]\,,\nonumber\\
E_3&=&-e_{\hat \theta}\,,
\end{eqnarray}
and
\beq
\tau_1=\gamma_\phi^2\nu^{\hat \phi}\frac{N}{r\gamma_K^2}+\nu^{\hat r}a(U)_{\hat \phi}\,.
\eeq

\noindent
{\bf  \underline{Case 3}:  Spatial acceleration, $a(U)_{\hat t}=0$}

In the special case $a(U)_{\hat t}=0$, namely $\nu^{\hat r}a(U)_{\hat r}+\nu^{\hat \phi}a(U)_{\hat \phi}=0$ we have
\begin{eqnarray}
\label{fs_case3}
E_1 &=&\frac{1}{\kappa }\left(a(U)^{\hat r}e_{\hat r}+a(U)^{\hat \phi}e_{\hat \phi}\right)
=\frac{a(U)^{\hat \phi}}{\kappa \nu^{\hat r}}(-\nu^{\hat \phi}e_{\hat r}+\nu^{\hat r}e_{\hat \phi})\,,\nonumber\\
E_2&=&-\frac{\gamma}{\nu}\left[
\nu^2n+\nu^{\hat r}e_{\hat r}+\nu^{\hat\phi}e_{\hat \phi}
\right]\,,\nonumber\\
E_3&=&-e_{\hat \theta}\,,
\end{eqnarray}
where $\nu=\sqrt{\nu_{\hat r}^2+\nu_{\hat \phi}^2}$ and
\beq
\fl\qquad
\kappa=\sqrt{a(U)_{\hat r}^2+a(U)_{\hat \phi}^2}=\frac{\nu}{|\nu_{\hat r}|}|a(U)_{\hat \phi}|\,,\qquad
\tau_1=\frac{N\nu_K^2}{r}\frac{\nu^{\hat \phi}}{\nu^2}+ \frac{a(U)_{\hat \phi}}{\nu^{\hat r}}\,.
\eeq

Evidently in the special cases here considered we have not specified the physical source of the acceleration nor the type of orbit. In what follows we shall discuss explicit 
examples.

\section{Explicit examples}

\subsection{Charged particles in external magnetic fields}

The solution of an electromagnetic field added to the Schwarzschild background has been found by Bi\v c\'ak and Janis \cite{bicak} in 1985 and it is presented in the appendix.
Here we limit our consideration to an electromagnetic field $F$ stemming from a vector potential $A$ given by 
\beq
\label{em_field}
\fl
A\equiv A_{(0)}=\frac12 B_0 r^2\sin^2\theta \rmd\phi\,, \quad F=dA=B_0r \sin\theta  (\sin\theta \rmd r+r \cos\theta \rmd \theta )\wedge \rmd \phi\,,
\eeq
where $B_0$ is an arbitrary constant.

The electric and magnetic fields relative to a given observer $u$ are generally defined as
\beq
E(u)=F\rightcontract u\,, \qquad B(u)={}^*F\rightcontract u \,,
\eeq
where $\rightcontract$ denotes right contraction and ${}^*$ the spacetime duality operation, specifically
\beq
{}^*F_{\alpha\beta}=\frac12 \eta_{\alpha\beta\gamma\delta} F^{\gamma\delta}\,.
\eeq
It should be kept in mind that the components of the electric field always appear with the specification of the observer; therefore they should not be confused with the legs of a tetrad.

With respect to a static observer $u=n$  the electromagnetic field (\ref{em_field}) is a purely magnetic one, i.e. the electric field vanishes ($E(n)=0$) and 
\begin{eqnarray}
\label{Bzamo}
B(n)&=& B_0 [-\cos\theta e_{\hat r}+N\sin \theta e_{\hat \theta}]\,.
\end{eqnarray}
On the equatorial plane $\theta=\pi/2$, $B(n)$ reduces to
\begin{eqnarray}
\label{Bzamo2}
B(n)&=& B_0 N e_{\hat \theta}\,.
\end{eqnarray}
We recall that we only consider motion in the equatorial plane $\theta=\pi/2$.

The force on a charged test particle $U$  (with mass $m$ and charge $q$)   due to the given electromagnetic field is 
\beq
\fl\quad
f_{\rm (em)}(U)_{\hat \alpha}=qE(U)_{\hat \alpha}=qF_{\hat \alpha \hat \beta}U^{\hat \beta}=
q\gamma [\nu \times B(n)]_{\hat \alpha}=
\gamma q\eta(n)_{\hat \alpha \hat \beta \hat \gamma}\nu^{\hat \beta} B(n)^{\hat \gamma}\,,
\eeq
where $\eta(n)_{\hat \alpha \hat \beta \hat \gamma}=\eta_{\hat t \hat \alpha \hat \beta \hat \gamma}$.
The components of the electric field $E(U)$ are
\begin{eqnarray}
\label{fcomps}
\fl\quad
E(U)_{\hat r}=-\gamma \nu^{\hat \phi} B(n)^{\hat \theta} \,,  \quad
E(U)_{\hat \theta}=0\,,\quad
E(U)_{\hat \phi}=\gamma \nu^{\hat r} B(n)^{\hat \theta} \,,
\end{eqnarray}
so that 
\beq
\fl\qquad
f_{\rm (em)}(U)=q \gamma N B_0 [-\nu^{\hat \phi} e_{\hat r}+\nu^{\hat r} e_{\hat \phi}]= m \zeta_0 \gamma N [-\nu^{\hat \phi} e_{\hat r}+\nu^{\hat r} e_{\hat \phi}]\,,
\eeq
where we have introduced for convenience the \lq\lq cyclotron frequency"
\beq
\zeta_0=\frac{qB_0}{m}\,.
\eeq
The motion of the  particle under the effect of this force is governed by 
\beq
m a(U)=f_{\rm (em)}(U)\,,
\eeq
and since $f_{\rm (em)}(U)_{\hat t}=0$ it is described by Case 3 of the previous section.
The equations of motion are then, from Eqs. (\ref{eq_nus})
\begin{eqnarray}
\label{equatorial_CM}
\frac{\rmd \nu^{\hat r}}{\rmd \tau}&=& N \left\{ -\zeta_0 \nu^{\hat \phi}+\frac{\gamma}{r}[\nu^{\hat \phi}{}^2-\nu_K^2(1-\nu^{\hat r}{}^2)]\right\}\,,\nonumber \\
\frac{\rmd \nu^{\hat \phi}}{\rmd \tau}&=& -N\nu^{\hat r}\left(-\zeta_0 +\frac{\gamma}{r\gamma_K^2}\nu^{\hat \phi}\right)\,,
\end{eqnarray}
together with 
\begin{eqnarray}
\label{equatorial_CM2}
\frac{\rmd r}{\rmd \tau}&=& \gamma N \nu^{\hat r}\,,\qquad
\frac{\rmd \phi}{\rmd \tau}=\frac{\gamma}{r} \nu^{\hat \phi}\,.
\end{eqnarray}

We then proceed to the analysis of these orbits, considering their intrinsic properties, the equilibrium condition and its stability and the actual form of the orbit obtained by a direct numerical integration of Eqs. (\ref{equatorial_CM}) and (\ref{equatorial_CM2}).

\subsubsection{Intrinsic analysis of the orbits}

A Frenet-Serret frame adapted to the particle orbit $U$ and satisfying Eq. (\ref{FSeqs}) is given, from Eq. (\ref{fs_case3}), by
\begin{eqnarray}
\fl\quad
E_1=\frac{1}{\nu}(-\nu^{\hat \phi}e_{\hat r}+\nu^{\hat r}e_{\hat \phi})\,, \qquad
E_2=\frac{\gamma}{\nu}(\nu^2n+\nu^{\hat r}e_{\hat r}+\nu^{\hat \phi}e_{\hat \phi})\,, \qquad
E_3=-e_{\hat\theta}\,, 
\end{eqnarray}
modulo a sign choice.
The magnitude of the acceleration $\kappa$ and the first torsion $\tau_1$ turn out to be
\begin{eqnarray}
\fl\qquad
\kappa=\frac{1}{m}\, ||f_{\rm (em)}(U)||
=\gamma N\zeta_0\nu\,, \qquad
\tau_1=-\gamma N\zeta_0 -\frac{M}{Nr^2\nu^2}\nu^{\hat \phi}\,. 
\end{eqnarray}

\subsubsection{Equilibrium solution and stability}

The charged particle $U$ moves under the combined effect of the background metric and the external electromagnetic field.
When these two effects balance each other then we have an equilibrium solution for the equatorial orbit.

For the system (\ref{equatorial_CM}) and (\ref{equatorial_CM2}) the equilibrium solution is associated with a spatially circular orbit such that $\nu^{\hat r}=0$ and
\beq
\label{equiCM}
-\zeta_0 \nu^{\hat \phi}+\frac{\gamma}{r}(\nu^{\hat \phi}{}^2-\nu_K^2)=0\,,
\eeq
with $\gamma=(1-\nu^{\hat \phi}{}^2)^{-1/2}$.
Hereafter we shall assume $B_0>0$ without any loss of generality; for particles moving in the equatorial plane of the Schwarzschild spacetime this means a magnetic field aligned with the negative $z$ direction (i.e. aligned with $\partial_\theta$).
 
Eq. (\ref{equiCM}) determines  the equilibrium radius $r_0$
in terms of the orbital speed $\nu^{\hat \phi}=\pm\nu_0$ ($\nu_0>0$) at that radius (once $\nu_K$ is re-expressed in terms of $r_0$, namely $\nu_K=\sqrt{{M}/{(r_0-2M)}}$).
A solution for $r_0$ is easily found if we introduce  the rapidity parameter  $\nu_0=\tanh \alpha$. In fact, Eq. (\ref{equiCM}) becomes
\beq
\label{eqil_CMgen}
(r_0-3M)\sinh^2\alpha \pm r_0\zeta_0 (r_0-2M)\sinh \alpha +M=0\,,
\eeq
and it can be solved for $\sinh \alpha$. 
In the limiting case of a flat spacetime ($M=0$) the above relation reduces to
\beq
\sinh \alpha (\sinh\alpha \mp r_0\zeta_0)=0\,,
\eeq
and the corresponding solutions in terms of $\nu_0$ are given by
\beq
\label{nr_lim_eq}
\nu_0=0\,,\qquad \nu_0=\frac{r_0 |\zeta_0|}{\sqrt{1+r_0^2 \zeta_0^2}}\,.
\eeq
In the nonrelativistic limit of $r_0 \zeta_0 \ll 1$ Eq. (\ref{nr_lim_eq}) implies the known result
\beq
 \nu_0= r_0 |\zeta_0| \,.
\eeq
Note that in the weak-field  slow-motion limit ($M/r_0\ll 1$, $\nu_0 \ll 1$) the equilibrium condition (\ref{eqil_CMgen}) implies 
\beq
\label{eq_CM}
-\zeta_0 \nu_0 = \frac{M}{r_0^2}-\frac{\nu_0^2}{r_0}
\eeq
or, more explicitly, the balance between Lorentz, gravitational and centripetal forces
\beq
\frac{m\nu_0^2}{r_0} = \frac{m M}{r_0^2}-q B_0  \nu_0\,.
\eeq

\begin{figure} 
\begin{center}
\includegraphics[scale=.3]{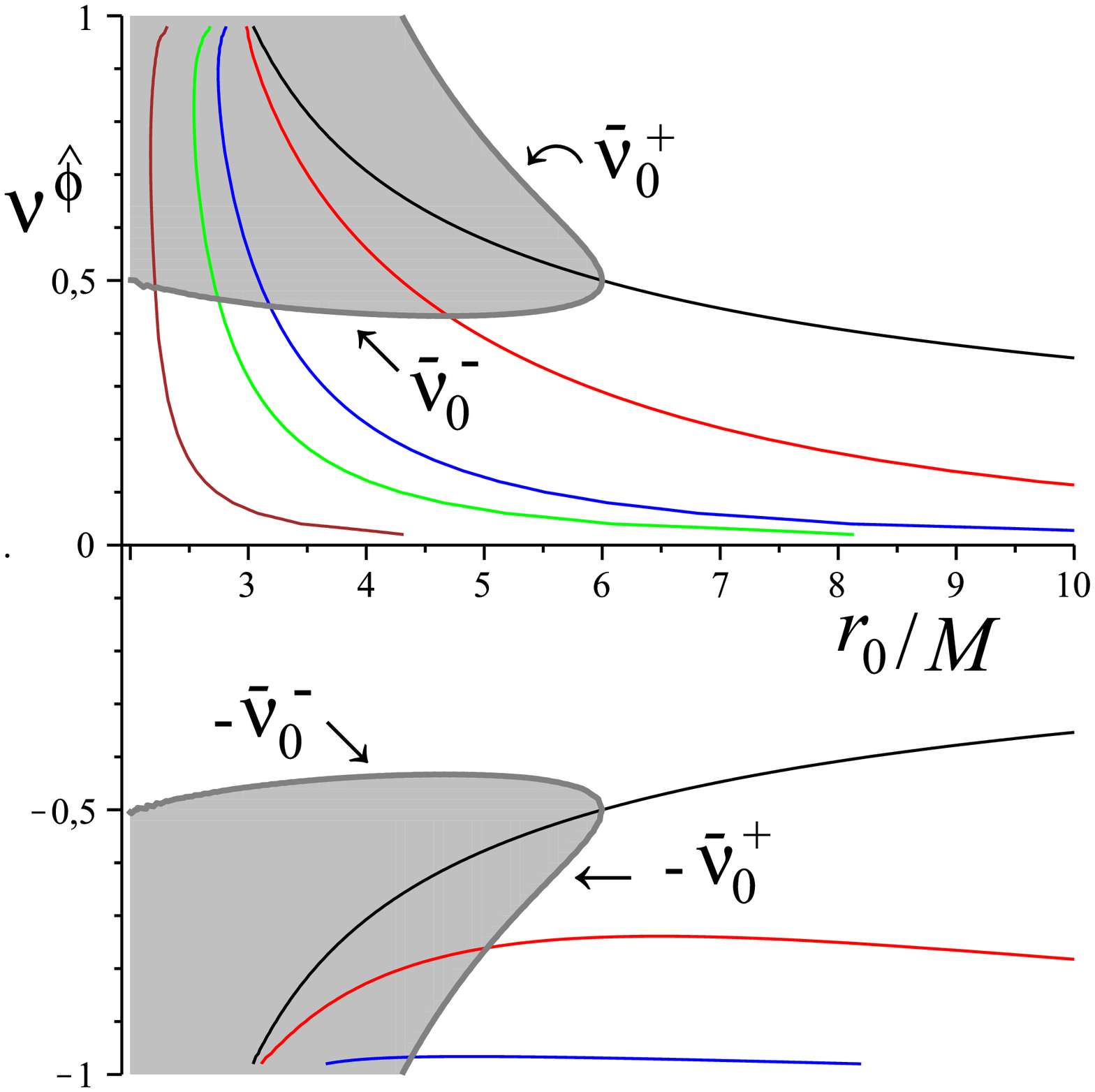}
\end{center}
\caption{
Charged particle in an external magnetic field. 
The equilibrium azimuthal velocity $\nu^{\hat\phi}$ is plotted as a function of $r_0/M$ for fixed values of $M\zeta_0=[0\, {\rm (black)}, -0.1\, {\rm (red)}, -0.5\, {\rm (blue)}, -1\, {\rm (green)}, -5\, {\rm (brown)}]$.
The corresponding equilibrium orbits are stable outside the shaded region.
For every fixed value of the equilibrium radius $r_0/M$ there exist in general two values of the azimuthal velocity corresponding to co-rotating and counter-rotating orbits.
} 
\label{fig:equil_magn}
\end{figure}

The stability of the  equilibrium orbit  can be studied by looking for solution of the perturbative problem
\beq\fl\qquad
r=r_0 +r_1(\tau)\,,\quad \phi=\phi_0(\tau)+\phi_1(\tau)\,,\quad \nu^{\hat r}=\nu^{\hat r}_1(\tau)\,,\quad \nu^{\hat \phi}=\pm\nu_0+\nu^{\hat \phi}_1(\tau)\,.
\eeq
From the general equations of motion (\ref{equatorial_CM}) and (\ref{equatorial_CM2}), the first order perturbations satisfy the following equations
\begin{eqnarray}
\fl\quad
\frac{\rmd r_1}{\rmd \tau}&=& \gamma_0 N \nu^{\hat r}_1\,, \nonumber\\
\fl\quad
\frac{\rmd \phi_1}{\rmd \tau}&=& -\frac{\gamma_0}{r_0}\left[\pm\nu_0\frac{r_1}{r_0}-\gamma_0^2 \nu^{\hat \phi}_1\right]\,, \nonumber\\
\fl\quad
\frac{\rmd \nu^{\hat r}_1}{\rmd \tau}&=& -\frac{\gamma_0N}{r_0}\left\{[\nu_0^2-\nu_K^2(1+\nu_K^2)]\frac{r_1}{r_0}-\frac{\gamma_0^2}{(\pm\nu_0)} [\nu_0^2+\nu_K^2-2\nu_0^2\nu_K^2]\nu^{\hat \phi}_1\right\}\,, \nonumber\\
\fl\quad
\frac{\rmd \nu^{\hat \phi}_1}{\rmd \tau}&=&-\frac{N\nu_K^2}{r_0\gamma_0(\pm\nu_0)}\nu^{\hat r}_1\,,
\end{eqnarray}
that is, formally, a linear system of the type
\beq
\frac{dX^\alpha}{d\tau}=A^\alpha{}_\beta X^\beta
\eeq
where $X=[ r_1,\phi_1,\nu^{\hat r}_1,\nu^{\hat \phi}_1]$.
The eigenvalues associated with the stability matrix $A$ are $\lambda_1=0=\lambda_2$, $\lambda_3=-\lambda_4=i\Lambda$, with
\beq
\label{Lambda}
\Lambda
=\frac{\gamma_0N}{r_0|\nu_0|}\sqrt{(\nu_0^2-\nu_K^2)^2+\nu_0^2\nu_K^2(1-4\nu_K^2)}\,,
\eeq
for $r_0>6M$, which vanishes at 
\beq
\nu_0=\bar \nu_0^\pm= \nu_K \left[\frac12 +2\nu_K^2 \pm \frac12 \sqrt{(4\nu_K^2-1)(4\nu_K^2+3)}  \right]^{1/2}\,.
\eeq
Therefore, the solution of the perturbed system exhibits an oscillating behavior with proper frequency $\Lambda$ for $r_0>6M$, which assures the existence of a stability regime in that region.
In fact, the second term in the square root in Eq. (\ref{Lambda}) is always positive there.
It vanishes at $r_0=6M$ (where $\bar \nu_0^+=\bar \nu_0^-$), so that the eigenvalues all vanish for the circular geodesics and $\nu_0=1/2=\nu_K|_{r_0=6M}$.
For $r_0<6M$ instead an instability region appears (shaded area in Fig. \ref{fig:equil_magn}) corresponding to negative values of the argument of the square root in Eq. (\ref{Lambda}), implying that the nonvanishing eigenvalues are both real and of different sign, i.e. $\lambda_1=0=\lambda_2$, $\lambda_3=-\lambda_4=-|\Lambda|$.

\begin{figure} 
\begin{center}
\includegraphics[scale=.35]{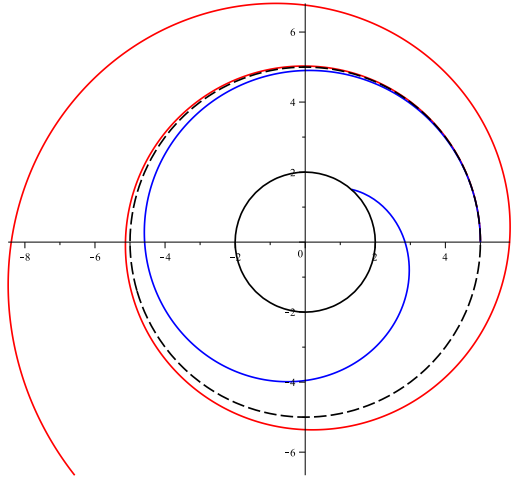}
\end{center}
\caption{
Charged particle in an external magnetic field. 
The orbit corresponding to $M\zeta_0=0.01$ is shown for the following choices of initial conditions:
$r(0)=5M$, $\phi(0)=0$, $\nu^{\hat r}(0)=0$, and $\nu^{\hat \phi}(0)=[0.59\, {\rm (blue)},0.598\, {\rm (black)},0.6\, {\rm (red)}]$. 
For the selected values of $r(0)/M$ and $M\zeta_0$ there exist two equilibrium values $\nu^{\hat \phi}(0)=\nu_0^-\approx-0.557$, $\nu^{\hat \phi}(0)=\nu_0^+\approx0.598$. 
Therefore, for values of $\nu^{\hat \phi}(0)$ close to the equilibrium value $\nu_0^+$ shown in the plot the orbit spirals either inward or outward. 
The inner circle represents the black hole horizon at $r=2M$.
} 
\label{fig:1}
\end{figure}

\begin{figure} 
\begin{center}
\[
\begin{array}{cc}
\includegraphics[scale=.25]{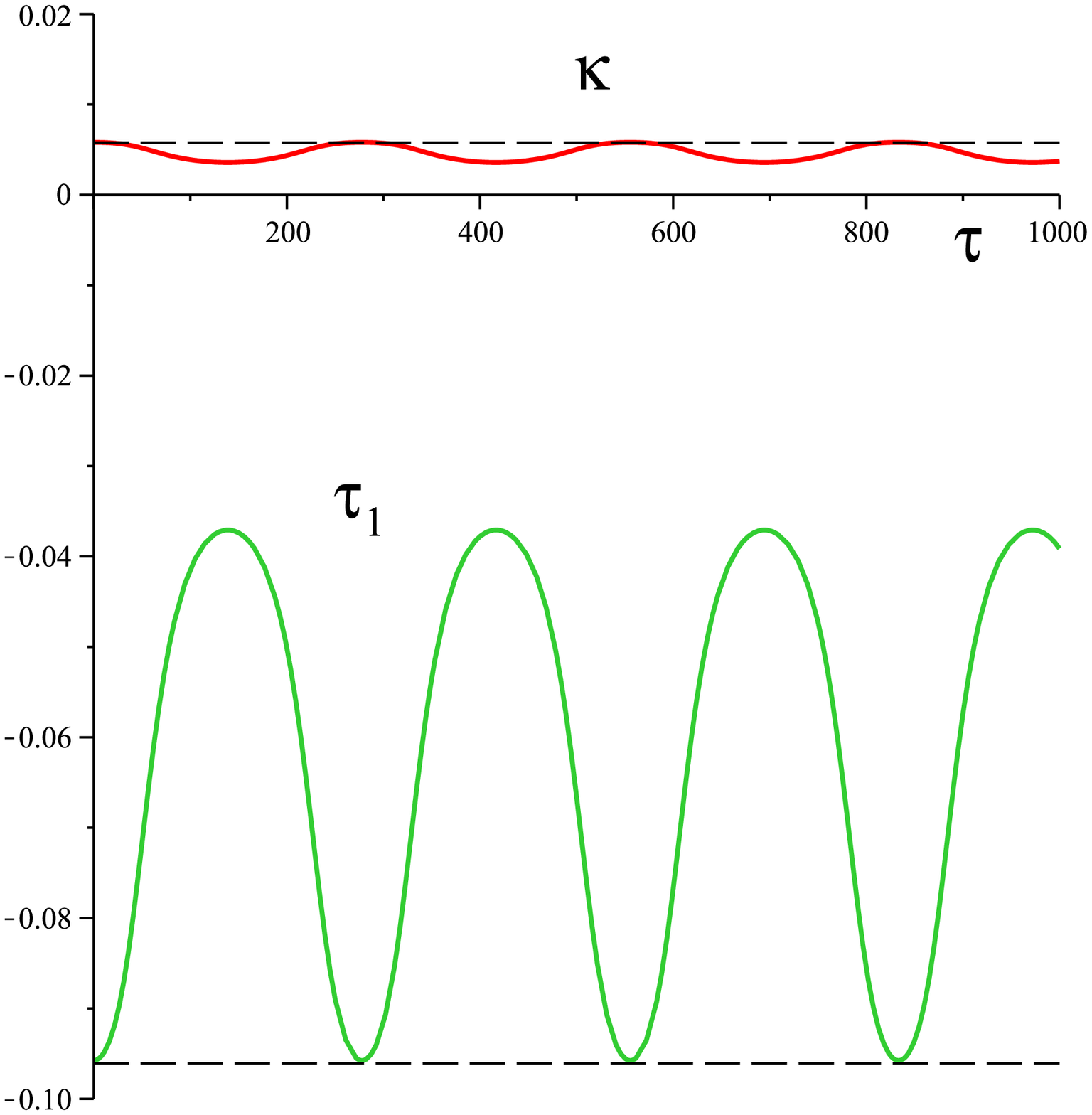}&
\includegraphics[scale=.25]{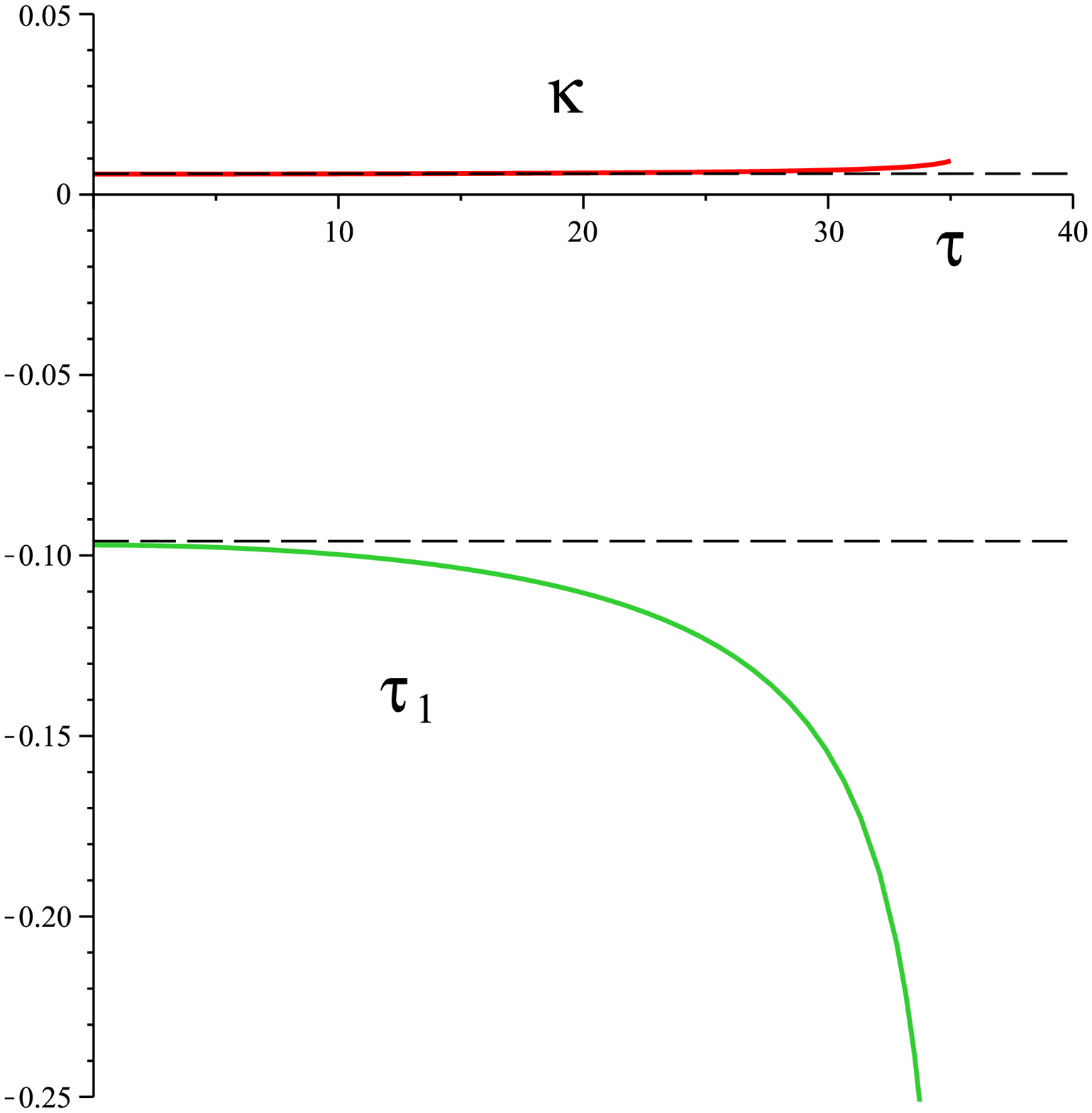}\\[.5cm]
(a)&(b)\\[.5cm]
\end{array}
\]
\end{center}
\caption{
Charged particle in an external magnetic field. 
The behaviours of the signed magnitude $\kappa$ of the 4-acceleration and the first torsion $\tau_1$ are shown as functions of the proper time $\tau$ for the same choice of parameters and initial conditions as in Fig. \ref{fig:1}. 
Figure (a) corresponds to the orbit spiraling outwards, whereas Fig. (b) to that falling down to the hole.
The dashed lines correspond to the constant values $\kappa\approx0.006$ and $\tau_1\approx0.096$ of the equilibrium solution.    
} 
\label{fig:2}
\end{figure}

Figure \ref{fig:1} shows typical orbits for a small value of the parameter $M\zeta_0$ and initial conditions close to the equilibrium solution.
The orbit spirals either inward or outward depending on whether the initial value of the azimuthal velocity is less or greater than the critical equilibrium one.
The corresponding behaviors of the signed magnitude of the acceleration and first torsion are shown in Fig. \ref{fig:2}.


\begin{figure} 
\begin{center}
\[
\begin{array}{cc}
\includegraphics[scale=.25]{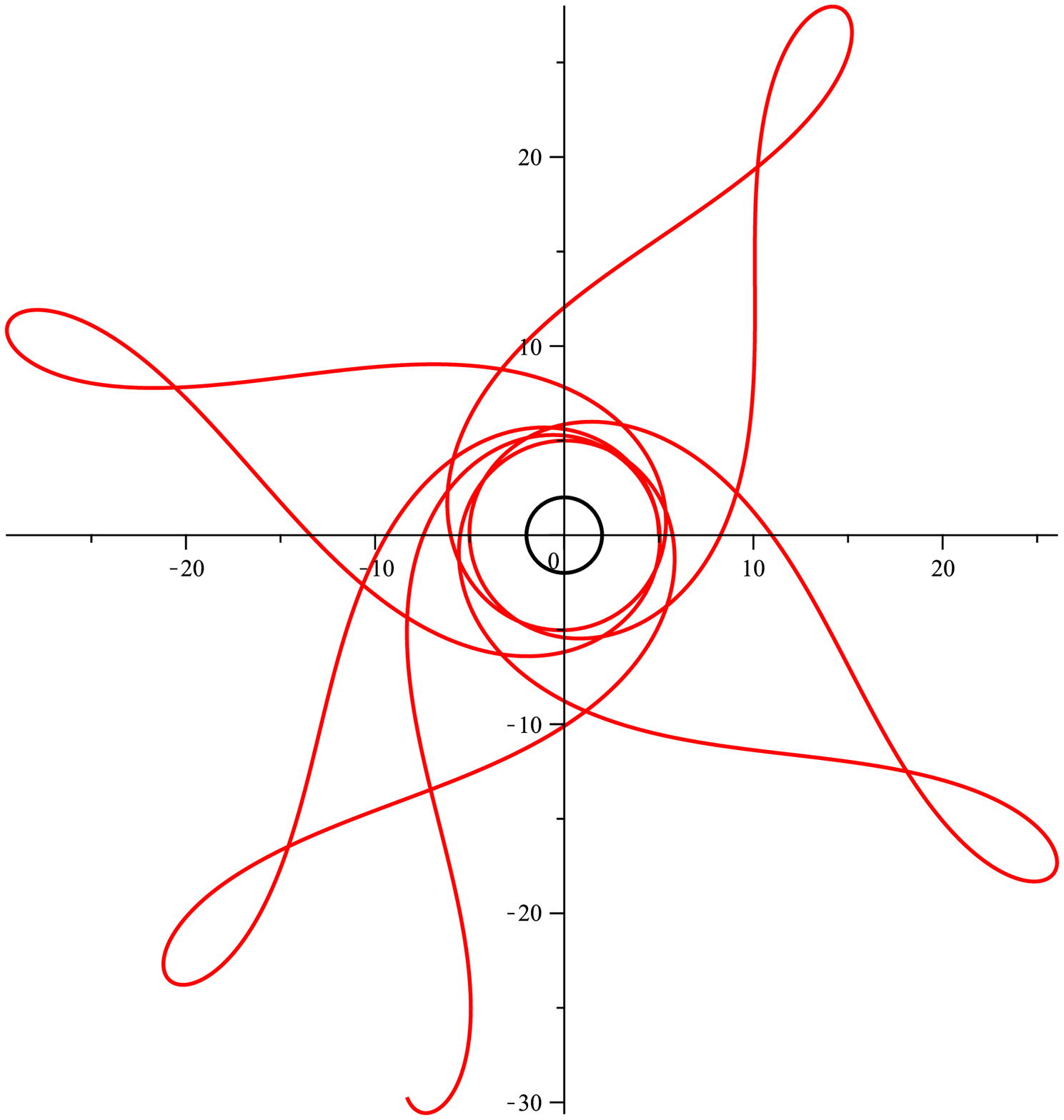}&
\includegraphics[scale=.25]{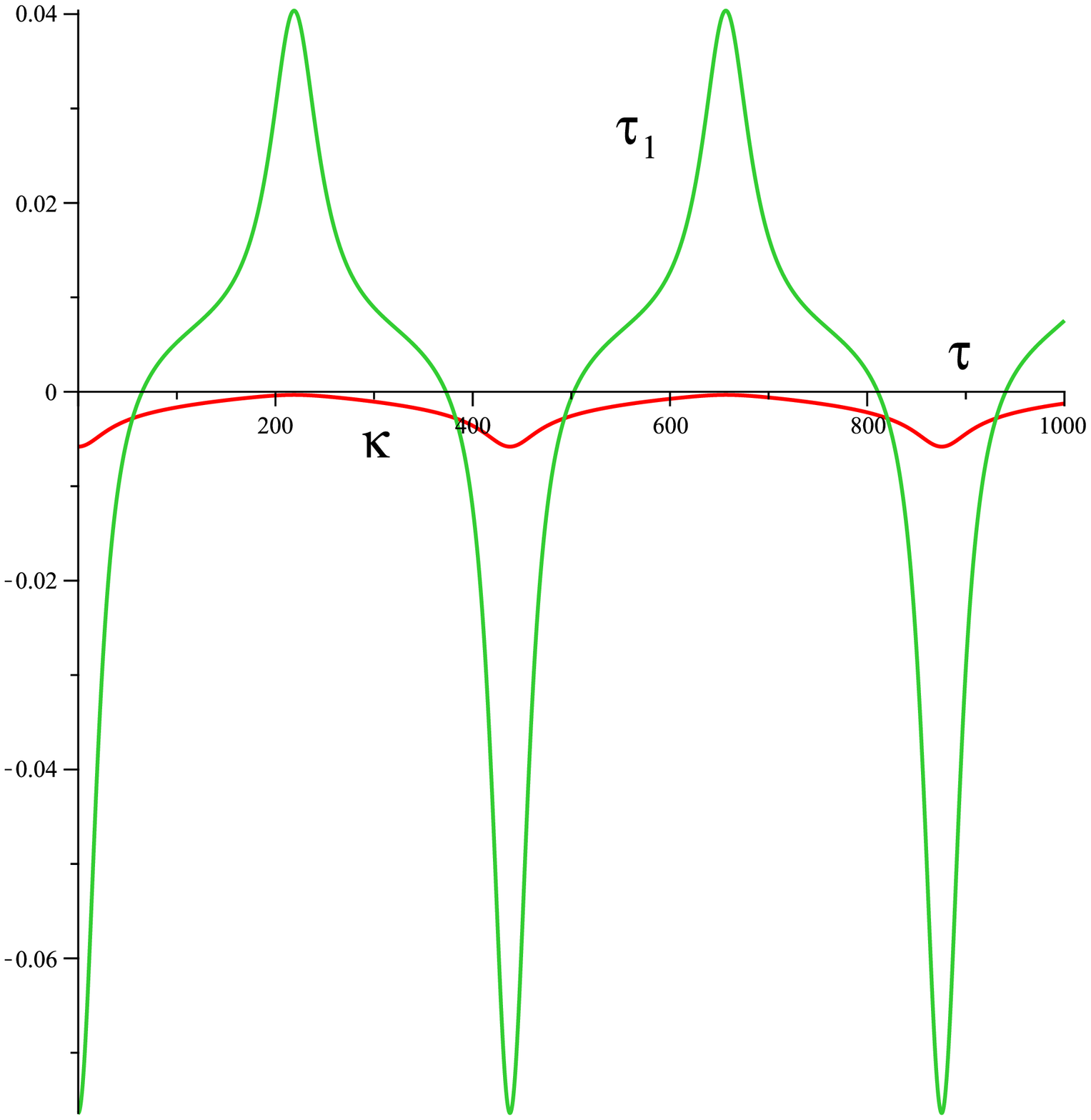}\\[.5cm]
(a)&(b)\\[.5cm]
\end{array}
\]
\end{center}
\caption{
Charged particle in an external magnetic field. 
The orbit corresponding to $M\zeta_0=-0.01$ is shown in Fig. (a) for the choice of initial conditions
$r(0)=5M$, $\phi(0)=0$, $\nu^{\hat r}(0)=0$, and $\nu^{\hat \phi}(0)=0.6$.  
The equilibrium values of the azimuthal velocity are $\nu^{\hat \phi}(0)\approx[-0.598,0.557]$.
Figure (b) shows the corresponding behaviors of the signed magnitude $\kappa$ and the first torsion $\tau_1$ as functions of the proper time $\tau$.        
} 
\label{fig:3}
\end{figure}

\begin{figure} 
\begin{center}
\[
\begin{array}{cc}
\includegraphics[scale=.25]{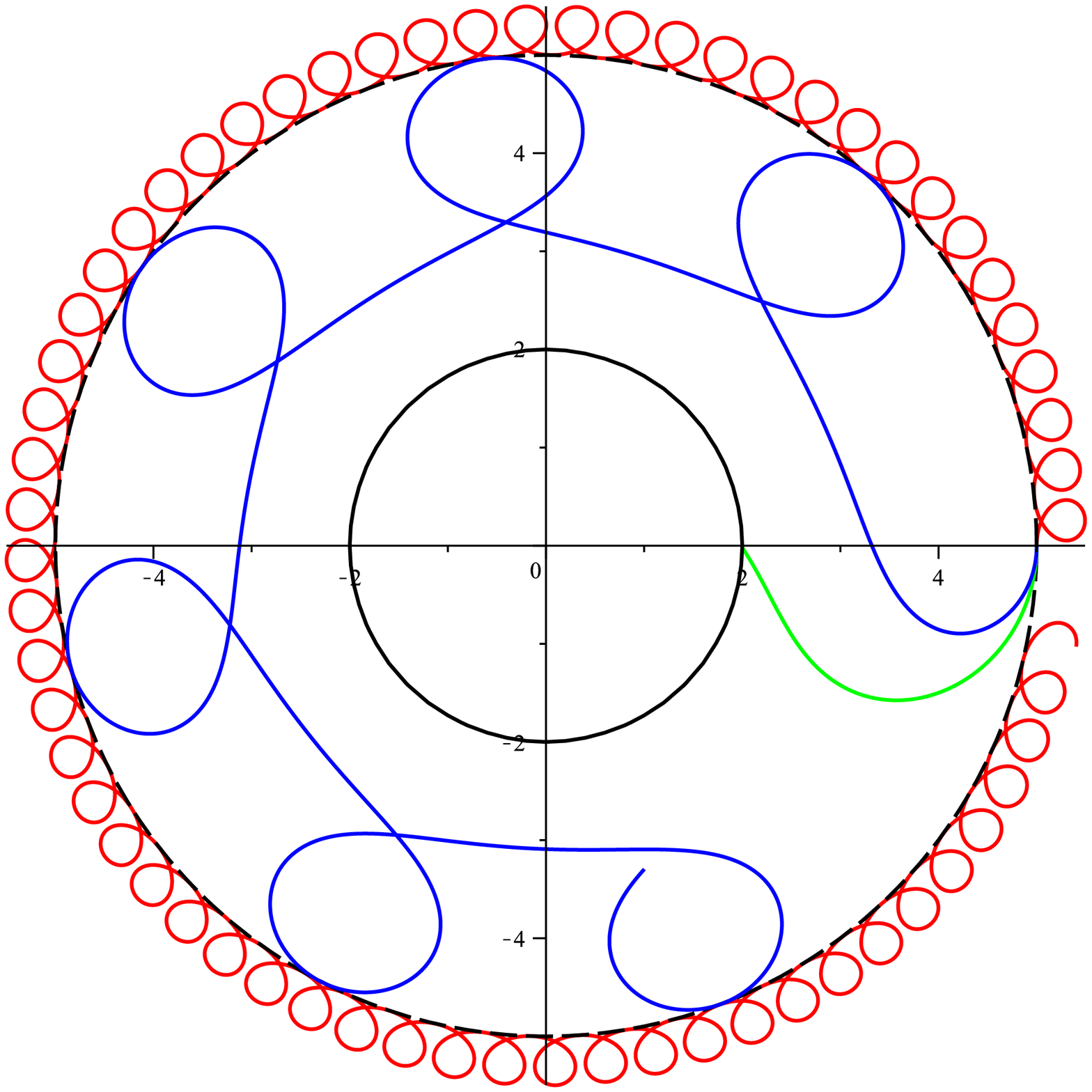}&
\includegraphics[scale=.25]{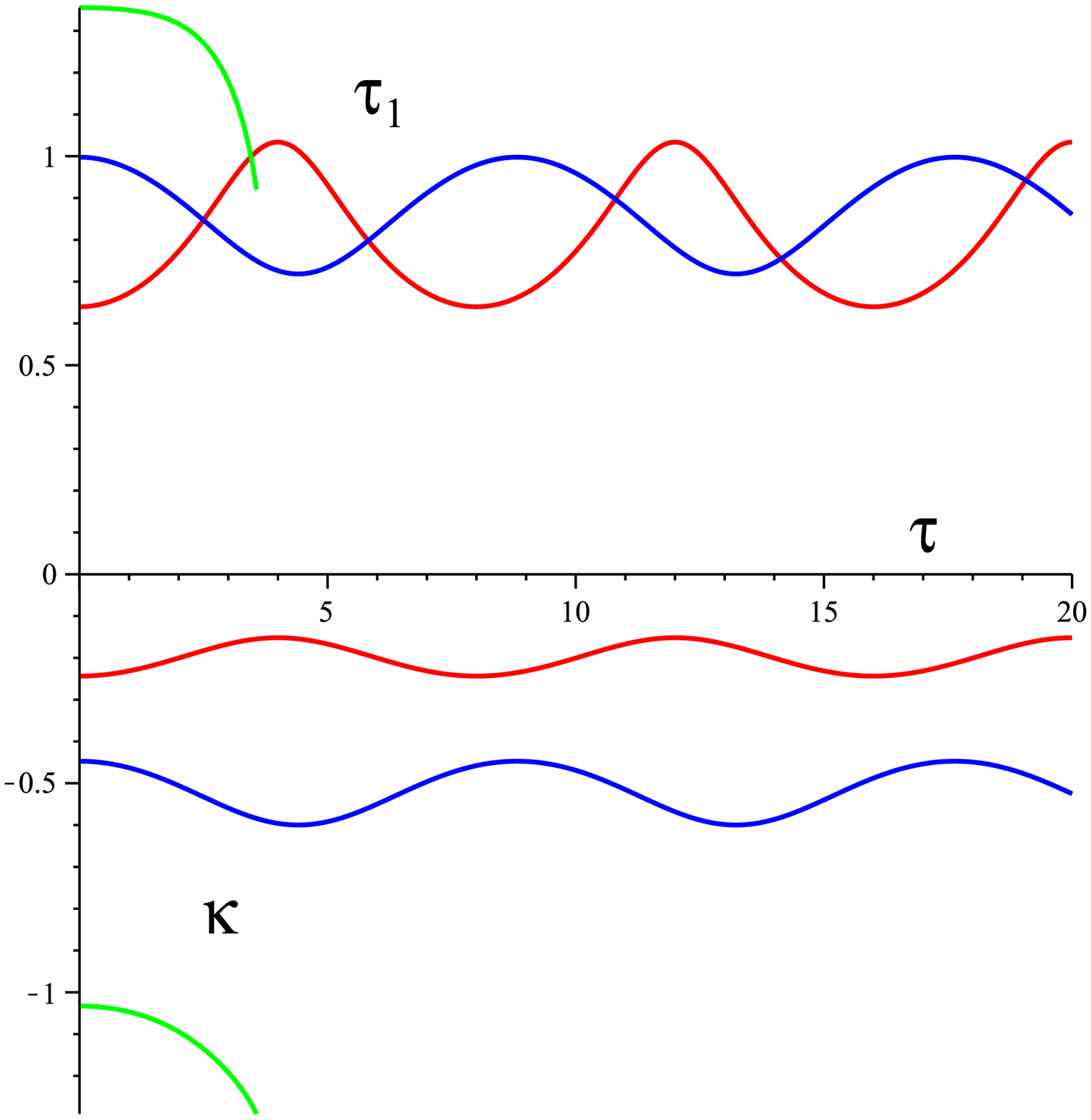}\\[.5cm]
(a)&(b)\\[.5cm]
\end{array}
\]
\end{center}
\caption{
Charged particle in an external magnetic field. 
The orbit corresponding to $M\zeta_0=-1$ is shown in Fig. (a) for the following choices of initial conditions:
$r(0)=5M$, $\phi(0)=0$, $\nu^{\hat r}(0)=0$, and $\nu^{\hat \phi}(0)=[-0.8\, {\rm (green)},-0.5\, {\rm (blue)},0.066\, {\rm (black)},0.3\, {\rm (red)}]$.  
For the selected values of $r(0)/M$ and $M\zeta_0$ the equilibrium values of the azimuthal velocity are $\nu^{\hat \phi}(0)\approx[-0.991,0.066]$.
Figure (b) shows the behaviors of the signed magnitude $\kappa$ and the first torsion $\tau_1$ as functions of the proper time $\tau$ for the same choice of parameters and initial conditions as in Fig. (a).   
The critical values corresponding to equilibrium are $\kappa\approx-0.051$ and $\tau_1\approx-0.007$.     
} 
\label{fig:4}
\end{figure}

Further examples of orbits with corresponding curvature and torsions are shown in Figs. \ref{fig:3} and \ref{fig:4}.
For every choice of $r(0)/M$ and $M\zeta_0$ there exist in general two equilibrium values $\nu^{\hat \phi}(0)=\nu_0^-$, $\nu^{\hat \phi}(0)=\nu_0^+$, where $r(0)$ and $\nu^{\hat \phi}(0)$ are the initial values of the radius and azimuthal speed at $\tau=0$.
This is evident 
from Fig. \ref{fig:equil_magn}, where the equilibrium azimuthal velocity $\nu^{\hat\phi}$ is plotted as a function of $r_0/M$ for fixed values of $M\zeta_0$ and the values $\nu_0^\pm$ can be located once the equilibrium radius is fixed.
As a general feature for $\nu^{\hat \phi}(0)<\nu_0^-$ and $\nu^{\hat \phi}(0)>\nu_0^+$ the orbits exhibit an oscillating behavior around a mean value $\bar r>r(0)$; for $\nu_0^-<\nu^{\hat \phi}(0)<\nu_0^+$ instead the oscillations are around a mean value $\bar r<r(0)$ which may cause the particle to fall down to the horizon.
The maximum amplitude in both cases is $|\bar r-r(0)|$.


\subsection{Spinning particles}

The motion of a test spinning particle in a gravitational field has been extensively investigated for the relevance of rotation among the astrophysical bodies. 
Limiting one's interest to the pole-dipole approximation, the corresponding equations of motion are given by the Mathisson-Papapetrou equations \cite{math37,papa51}. 
To become a close set, however, the above equations need supplementary conditions, specifically denoted as Papapetrou-Corinaldesi (PC) \cite{cori51}, Pirani (P) \cite{pir56} and Tulczyjew (T) \cite{tulc59}, each of one leading to different solutions. 
In a series of papers \cite{clockschw,clockkerr,clockweyl,bdfgj1,bdfgj2} the effects of each supplementary condition on equatorial orbits in various spacetime metrics have been studied and the main physically relevant conclusion was that the P and T conditions are the only relevant ones although not clear reason for selecting one instead of the other can be theoretically motivated. 
Most important is the result that, in the limit of small spin, the above two conditions are equivalent. Since these conditions are considered here, we assume them without further notice.

The motion of a test spinning body is driven by the force
\beq
f_{\rm (spin)}(U){}^{\mu}=-\frac12R^{\mu}{}_{\nu\alpha\beta}U^{\nu}S^{\alpha\beta}\,,
\eeq 
so that the equations of motion write as $m a(U)^\mu=f_{\rm (spin)}(U){}^{\mu}$. 
Here $U$ is the particle $4$-velocity and $S^{\mu\nu}$ denotes the (antisymmetric) spin tensor, conveniently represented in terms of the spin vector 
\beq
S^\beta=\frac12 U_\alpha \eta^{\alpha\beta\mu\nu}S_{\mu\nu}\,.
\eeq 

Let us recall that the particle is moving in the equatorial plane ($\theta=\pi/2$, $\nu^{\hat \theta}=0$)  and assume that its spin vector is constant and orthogonal to the motion plane, i.e.
\beq
S=-se_{\hat \theta}\,, 
\eeq
$s$ being the signed magnitude of the spin vector.
We will also use the spin dimensionless quantity 
\beq
{\hat s}=\frac{s}{mM}\,,
\eeq
which should be small in order to avoid back-reaction effects.

The nonvanishing components of the spin force are given by \cite{clockschw}
\begin{eqnarray}
f_{\rm (spin)}(U){}_{\hat t}&=&\frac{3M}{r^3}s\gamma^2\nu^{\hat \phi}\nu^{\hat r}\,, \qquad
f_{\rm (spin)}(U){}_{\hat r}=-\frac{3M}{r^3}s\gamma^2\nu^{\hat \phi}\,,
\end{eqnarray}
that is
\beq
\label{fspin}
f_{\rm (spin)}(U)=-\frac{3M}{r^3}s\gamma^2\nu^{\hat \phi}(\nu^{\hat r}n+e_{\hat r})\,,
\eeq
as in Case 1 of Section 3.

The equations of motion (\ref{eq_nus}) are then
\begin{eqnarray}
\label{equatorial_spin}
\frac{\rmd \nu^{\hat r}}{\rmd \tau}&=&-\frac{3M^2}{r^3}{\hat s}\frac{\gamma}{\gamma_r^2}\nu^{\hat \phi}+\frac{N\gamma}{r}[\nu^{\hat \phi}{}^2-\nu_K^2(1-\nu^{\hat r}{}^2)]\,, \nonumber\\
\frac{\rmd \nu^{\hat \phi}}{\rmd \tau}&=&\frac{3M^2}{r^3}{\hat s}\gamma\nu^{\hat \phi}{}^2\nu^{\hat r}-\frac{N\gamma}{\gamma_K^2r}\nu^{\hat \phi}\nu^{\hat r}\,.
\end{eqnarray}
Because  of their awkwardness, these  can be only studied numerically except for very special cases.
It results that for small values of $\hat s$ the resulting orbits spiral either inward or outward depending on whether the initial value of the azimuthal velocity is less or greater than the critical equilibrium one.
The behaviour is very similar to the one illustrated in Fig. \ref{fig:1}.

\subsubsection{Intrinsic analysis of the orbits}

A Frenet-Serret frame satisfying Eq. (\ref{FSeqs}) can be built up with the triad
\begin{eqnarray}
\fl\qquad
E_1=\gamma_r(\nu^{\hat r}n+e_{\hat r})\,, \qquad
E_2=\frac{\gamma_r}{\gamma}(e_{\hat\phi}+\gamma\nu^{\hat \phi}U)\,, \qquad
E_3=-e_{\hat\theta}\,, 
\end{eqnarray}
with associated curvature and first torsion  
\beq
\fl\qquad
\kappa= \frac{1}{m}\, ||f_{\rm (spin)}(U)||
=-\frac{3M^2}{r^3}{\hat s} \frac{\gamma^2}{\gamma_r}\nu^{\hat\phi}\,, \qquad
\tau_1= \gamma_r \nu^{\hat \phi}\left( \frac{\gamma_r N}{r}-\kappa \right)\,.
\eeq

\subsubsection{Equilibrium solution and stability}

An equilibrium circular orbit solution exists on the equatorial plane at $r=r_0$ only if $\nu^{\hat r}=0$ and $\nu^{\hat \phi}\equiv \pm\nu_0$, implying
\begin{eqnarray}
N(\nu_0^2-\nu_K^2)= \pm\frac{3M}{r_0^2}\frac{s}{m}\nu_0\,,
\end{eqnarray}
which can be easily solved for $\nu_0$.
In the weak-field and slow-motion limit the above relation admits the classical limit
\beq
\label{eq_spin}
m \frac{\nu_0^2}{r_0}= \frac{mM}{r_0^2}\pm \frac{3M}{r_0^3} s \nu_0\,.
\eeq
In this approximated situation one may compare Eq. (\ref{eq_spin}) with the equilibrium condition (\ref{eq_CM}) for a massive and charged particle subjected to a magnetic field; the equations coincide if  
\beq
|\zeta_0|=\left|\frac{qB_0}{m}\right| \quad \to \quad  
\frac{3M^2}{r_0^3}|{\hat s}|\,.
\eeq
This shows that in the case of a spinning particle the role of the dimensionless spin ${\hat s}$ is played by the electric charge.
Moreover, we clearly see that there exists a value of the average radius $r_0$ of the orbit for each \lq\lq equivalent" triplet $[q,B_0,m]$ specifying the particle for which the two cases coincide.

Finally, one may consider the stability of this orbit looking for solution of the perturbative problem
\beq
\fl\quad
r=r_0 +r_1(\tau)\,,\quad \phi=\phi_0(\tau)+\phi_1(\tau)\,,\quad \nu^{\hat r}=\nu^{\hat r}_1(\tau)\,,\quad \nu^{\hat \phi}=\pm\nu_0+\nu^{\hat \phi}_1(\tau)\,.
\eeq
The first order quantities satisfy the following system of equations
\begin{eqnarray}
\frac{\rmd r_1}{\rmd \tau}&=& \gamma_0 N \nu^{\hat r}_1\,, \nonumber\\
\frac{\rmd \phi_1}{\rmd \tau}&=& -\frac{\gamma_0}{r_0}\left[\pm\nu_0\frac{r_1}{r_0}-\gamma_0^2 \nu^{\hat \phi}_1\right]\,, \nonumber\\
\frac{\rmd \nu^{\hat r}_1}{\rmd \tau}&=& \frac{\gamma_0N}{r_0}\left\{[\nu_K^4-\nu_K^2(1-\nu_0^2)+2\nu_0^2]\frac{r_1}{r_0}+\frac{1}{(\pm\nu_0)} (\nu_0^2+\nu_K^2)\nu^{\hat \phi}_1\right\}\,, \nonumber\\
\frac{\rmd \nu^{\hat \phi}_1}{\rmd \tau}&=&-\frac{N(\pm\nu_0)}{r_0\gamma_0}\nu^{\hat r}_1\,.
\end{eqnarray}
The associated eigenvalues are $\lambda_1=0=\lambda_2$, $\lambda_3=-\lambda_4=\Lambda_s$, with
\begin{eqnarray}
\Lambda_s&=&\frac{\gamma_0N}{r_0}\sqrt{(\nu_0^2+\nu_K^2)^2+\nu_0^2-2\nu_K^2}\nonumber\\
&\equiv&\frac{\gamma_0N}{r_0}\sqrt{(\nu_0^2-\bar \nu_0^+{}^2)(\nu_0^2-\bar \nu_0^-{}^2)}\,,
\end{eqnarray}
where 
\beq
\bar \nu_0^\pm{}^2=-\nu_K^2-\frac12\pm\frac12\sqrt{1+12\nu_K^2}\,.
\eeq
Therefore, the equilibrium is stable only in the region wherein the argument of the square root is negative, i.e. for $-\bar \nu_0^+{}<\nu_0<\bar \nu_0^+$.
The solution of the perturbed system exhibits there an oscillating behavior with proper frequency $\Lambda_s$.

On the basis of the above analysis in the case of equilibrium solutions one cannot distinguish between the equatorial motion of a charged particle in an external electromagnetic field and a spinning neutral particle in the given background.


\begin{figure} 
\begin{center}
\includegraphics[scale=.3]{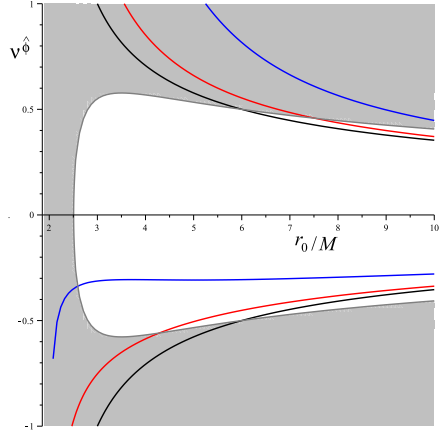}
\end{center}
\caption{
Spinning test particle. 
The equilibrium azimuthal velocity $\nu^{\hat\phi}$ is plotted as a function of $r_0/M$ for fixed values of ${\hat s}=[0\, {\rm (black)}, 1\, {\rm (red)}, 5\, {\rm (blue)}]$.
For every fixed value of the equilibrium radius $r_0/M$ there exist in general two values of the azimuthal velocity corresponding to co-rotating and counter-rotating orbits (intersections with solid curves in the plot).
The corresponding equilibrium orbits are stable outside the shaded region.
} 
\label{fig:equil_spin}
\end{figure}

\subsection{Particles with a magnetic dipole moment}

When to the Schwarzschild background is added a magnetic field as in Eq. (\ref{Bzamo2}) we can also discuss the motion of a magnetic dipole \cite{fdf_sorge,fdf_sorge_zilio,fdf_preti,fdf_preti2} with magnetic moment $\mu^\alpha=-(\mu/r)\delta^{\alpha}_{\theta}$.
The force driving the motion in this case is given by
\beq
f_{\rm (dip)}(U){}^{\sigma}=\eta^{\alpha\beta\gamma\delta}U_\alpha\mu_\beta\nabla^{\sigma}F_{\gamma\delta}\,,
\eeq
that is 
\beq
\label{fdip}
f_{\rm (dip)}(U)=\frac{2\mu M\gamma B_0}{r^2}(\nu^{\hat r}n+e_{\hat r})\,.
\eeq
This force is very similar to one experienced by a spinning particle (see Eq. (\ref{fspin}));
hence we only show numerical integration of the orbits.
In this case, the equilibrium solution $r=r_0$, $\nu=\nu_0$ is given by
\begin{eqnarray}
\label{eq_dipolo}
\beta\frac{M}{r_0}+N\gamma_0(\nu_0^2-\nu_K^2)=0\,,
\end{eqnarray}
where $\beta=2\mu B_0/m$.
In this  situation one may compare Eq. (\ref{eq_dipolo}) with  the corresponding equilibrium condition for a spinning particle (\ref{eq_spin}); the equations coincide if  
\beq
2\mu B_0=\mp \frac{3}{r_0}s\frac{\nu_0}{\sqrt{1-\nu_0^2}}\,.
\eeq
This shows that, at the equilibrium, one cannot distinguish between  a magnetic dipole moving in an external magnetic field   and a spinning particle 
subjected to gravitation only.

\subsection{Superposed magnetic field and motion of a spinning particle also endowed with a magnetic dipole}

In view of the various astrophysical applications it is also worth considering the combined effects of a magnetic dipole which is also spinning.
The force which drives the motion become then
\beq
\fl\quad
f_{\rm (ds)}(U)=f_{\rm (dip)}(U)+f_{\rm (spin)}(U)=\left(2\mu B_0 -3\frac{s\gamma \nu^{\hat \phi}}{r}\right)\frac{M\gamma}{r^2}(\nu^{\hat r}n+e_{\hat r})\,,
\eeq
where $f_{\rm (dip)}(U)$ is given by Eq. (\ref{fdip}) and $f_{\rm (spin)}(U)$ by Eq. (\ref{fspin}). The equations of motion are given by
 \begin{eqnarray}
\label{eq_nus2}
\frac{d \nu^{\hat r}}{d\tau}
&=& \frac{a(U)_{\hat r}}{\gamma\gamma_r^2}+
\frac{\gamma N}{r}[\nu^{\hat \phi}{}^2 -\nu_K^2 (1-\nu^{\hat r}{}^2)]\,,\nonumber \\
\frac{d  \nu^{\hat \phi}}{d\tau}
&=& -\nu^{\hat r}\nu^{\hat \phi} \left( 
\frac{a(U)_{\hat r}}{\gamma}
+\frac{\gamma N}{\gamma_K^2 r}\right)\,,
\end{eqnarray}
with
\beq
a(U)_{\hat r}=\left(2\mu B_0 -3\frac{s\gamma \nu^{\hat \phi}}{r}\right)\frac{M\gamma}{m r^2}\,.
\eeq
In this case, similarly to those discussed above, an equilibrium spatially circular orbits at $r=r_0$ exists, i.e. with $\nu^{\hat r}=0$ ($\gamma_r=1$) and $\nu^{\hat \phi}=$ const.,
such that 
\beq
\left(2\mu B_0 -3\frac{s\gamma \nu^{\hat \phi}}{r_0}\right)\frac{M\gamma}{m r_0^2}+\frac{\gamma^2 N}{r_0}[\nu^{\hat \phi}{}^2 -\nu_K^2]=0\,,
\eeq
that is
\beq
\left(\beta  -3\frac{\hat s M\gamma \nu^{\hat \phi}}{r_0}\right)\frac{M}{ r_0}+\gamma  N (\nu^{\hat \phi}{}^2 -\nu_K^2)=0\,.
\eeq
This equation is actually a polynomial relation in $\nu^{\hat \phi}$ of the $4^{th}$-order and can be solved exactly.
There is also the possibility for the azimuthal speed to keep the geodesic value, i.e. $\nu^{\hat \phi}=\nu_K$ at $r=r_0$ and then to behave as a neutral non-spinning body,  moving along a geodesic.
In this case the effects on the particle's motion by the magnetic dipole and the intrinsic spin with respect to a background gravitational field coupled with a uniform magnetic field, compensate each other so to annul their combined effects. The critical condition for \lq\lq hiding" the structures of the body
is given by
\beq
\label{beta}
\beta = 3\hat s\gamma_K \nu_K\frac{ M}{r_0}\,,
\eeq 
where, we recall, $\beta=2\mu B_0/m$, $\hat s=s/(mM)$ with $m$ and $M$ the masses of the orbiting body and of the source of the background field, $\mu$ is the body's magnetic dipole moment and $B_0$ the added external magnetic field.

Let us see whether Eq. (\ref{beta}) provides astrophysically plausible conditions. Let us first convert the latter equation into conventional units and denote the corresponding quantities by a tilde,
\beq\fl\quad
M=\frac{G \tilde M}{c^2}\,,\quad \nu=\frac{\tilde \nu}{c}\,,\quad \mu=\sqrt{\frac{G}{4\pi \epsilon_0 c^6}}\tilde \mu\,,\quad
B_0=\sqrt{\frac{4\pi \epsilon_0 G}{c^2}}\tilde B_0\,,\quad s=\frac{G}{c^3}\tilde s\,,
\eeq
so that Eq. (\ref{beta}) writes as
\beq
\tilde \mu \tilde B_0=\frac{3}{2r_0} \tilde s \gamma_K \tilde \nu_K \,.
\eeq
Recalling that $\tilde s \sim \tilde M R^2/T$, where $T$ is the period of rotation and that for a neutron star on average, we have $\tilde M\sim M_\odot\sim 2\cdot 10^{30}\, Kg$, $R\sim 10^4\,m$, $T\sim 1\, s$, $\tilde \nu_K \sim 5\cdot 10^6 \, m/s$ and, in the case of Sgr A$^*$, $r_0\sim 2\cdot 10^{13}\, m$, 
\beq
\tilde \mu \tilde B_0 \sim 1.5 \cdot 10^{32} \, Joule\,;
\eeq
hence with a value of $\tilde B_0\sim 1 \, Tesla$ (average interstellar magnetic field close to the black hole location) we find
$\tilde \mu \sim 10^{32} Joule/Tesla$, a value which is not too far from what is expected for a magnetized neutron star.
 
This circumstance makes it plausible and realistic a patent ambiguity in the observation of stellar fields around massive black holes.

\subsection{Poynting-Robertson effect}

Special attention has been given recently to the case of a radiation field  superposed to a Schwarzschild spacetime.  Scattering (absorbing and consequent re-emitting) of such radiation  by moving particles causes a  drag force which acts on the particles  determining deviations from geodesic motion termed as Poynting-Robertson effect.
Details can be found in Refs. \cite{PR1,PR2}  so we give here only a brief account. 

A (null) radiation field  added to a Schwarzschild spacetime is described by the energy-momentum tensor
\beq
T=\Phi^2 k \otimes k\,,\qquad 
k^\alpha k_\alpha=0\,, \qquad
k^\alpha \nabla_\alpha k^\beta=0\,, 
\eeq
where $k$ is a null and geodesic vector of the background  while the flux $\Phi$ is determined by $\nabla_\beta T^{\alpha\beta}=0$.
The force acting on a massive particle with $4$-velocity $U$  can be written as
\beq
f_{\rm (rad)}{}_\alpha=-\tilde\sigma P(U)_{\alpha \beta} T^\beta{}_\mu U^\mu\,,
\eeq 
where $P(U)=g+U\otimes U$ projects orthogonally to $U$ and $\tilde\sigma$ is a coefficient modeling the absorption and consequent re-emission of radiation by the particle.
Let us consider photons moving on the equatorial plane of the Schwarzschild spacetime, i.e. with momentum $k$ given by 
\beq
k={\mathcal E}(n)(n+\hat \nu_k)\,,\qquad \hat \nu_k \cdot \hat \nu_k=1\,,
\eeq
with ${\mathcal E}(n)$ being the energy of the photons relative to ZAMOS and $\hat \nu_k$ being their (spacelike, unitary) direction of propagation
\beq
\label{eq:phot}
\hat \nu_k=\sin \beta\, e_{\hat r}+\cos \beta\, e_{\hat \phi}\,.
\eeq
The photon energy as measured by an observer comoving with the particle given by
\beq
\fl\qquad
\label{E_di_U}
{\mathcal E}(U)\equiv -U\cdot k=\gamma {\mathcal E}(n)(1-\hat \nu_k \cdot \nu)
=\gamma {\mathcal E}(n)(1-\sin \beta\nu^{\hat r}-\cos \beta\nu^{\hat \phi})\,.
\eeq
whereas the relative energy of the photons with respect to ZAMOs is
\beq
\label{E_di_n}
{\mathcal E}(n)= - k \cdot n 
=\frac{{\mathcal E}}{N}\,,\qquad 
\eeq
Here 
${\mathcal E}=-k_t>0$ is a constant of the motion representing the conserved energy associated with the timelike Killing vector field; ${\mathcal L}=k_\phi$ is another constant of the motion representing the conserved angular momentum associated with the rotational Killing vector field. We use the notation
\beq
b\equiv\frac{{\mathcal L}}{{\mathcal E}}\,,
\eeq
for the photon impact parameter \cite{MTW} so that
\beq
\label{cosbeta}
\cos \beta =\frac{bN}{r} \quad \to \quad  
N|b\tan\beta| =\sqrt{r^2-b^2N^2}\,.
\eeq
We will restrict ourselves to the case of  photons with ${\mathcal E}>0$  so that ${\mathcal E}(n)>0$ and $k$ is a future-directed vector. 

The case $\sin \beta >0$ corresponds to outgoing photons (increasing $r$) and $\sin \beta <0$ to ingoing photons (decreasing $r$). 
The case $\sin\beta=0$ for spatially circular geodesic motion of the photons can only take place at $r=3M$, so we exclude it.

Since $k$ is completely determined,  the coordinate dependence of the quantity $\Phi$  follows 
from the conservation equations  $\nabla_\beta T^{\alpha\beta}=0$. The result is the following \cite{PR2}
\beq
\label{eq_phi}
\Phi^2=\frac{\Phi_0^2}{r N|b \tan \beta|}=\frac{\Phi_0^2}{r\sqrt{r^2-b^2N^2}} \,,
\eeq
where $\Phi_0$ is a constant.

The motion of a massive particle under the effect of this force is governed by the equations
\beq
m a(U)=f_{\rm (rad)}(U)\,,
\eeq
and apart from very special situations the analysis could  only be performed numerically.
However, since $f_{\rm (rad)}(U)\cdot U=0$ we have
\beq
f_{\rm (rad)}(U)^{\hat t}=\nu_{\hat r} f_{\rm (rad)}(U)^{\hat r}+\nu_{\hat \phi}f_{\rm (rad)}(U)^{\hat \phi}
\eeq
with 
\begin{eqnarray}
\fl\qquad
f_{\rm (rad)}(U)^{\hat r}&=&\sigma \Phi^2 {\mathcal E}(U) {\mathcal E}(n)[\sin \beta-\gamma^2 (1-\sin \beta\nu^{\hat r}-\cos \beta\nu^{\hat \phi}) \nu^{\hat r}]\,, \nonumber \\
\fl\qquad
f_{\rm (rad)}(U)^{\hat \phi}&=&\sigma \Phi^2 {\mathcal E}(U) {\mathcal E}(n)[\cos \beta-\gamma^2 (1-\sin \beta\nu^{\hat r}-\cos \beta\nu^{\hat \phi})\nu^{\hat \phi}] \,.
\end{eqnarray}
If we make explicit $\Phi$ as in Eq. (\ref{eq_phi})  and ${\mathcal E}(U)$ and ${\mathcal E}(n)$ as in Eqs. (\ref{E_di_U}) and (\ref{E_di_n}) the expression of the force depends on the single constant $m A\equiv\sigma \Phi_0^2 {\mathcal E}^2$.

\subsubsection{Equilibrium solution}

The general equations of motion (\ref{eq_nus}) admit in this case  an equilibrium solution at a fixed radius $r=r_0$ with
\beq
\nu^{\hat r}=0\,,\quad \nu^{\hat \phi}=\pm \nu_0\,,\quad \gamma=1/\sqrt{1-\nu_0^2}\,;
\eeq
in fact, recalling that $\nu^{\hat \phi}=$ const. at the equilibrium,  Eqs. (\ref{eq_nus})  simplify as 
\begin{eqnarray}
\label{moto_gen2e}
\frac{1}{m}f_{\rm (rad)}^{\hat r}&=&\frac{1}{m}[\sigma \Phi^2 {\mathcal E}(U) {\mathcal E}(n) \sin \beta_0]=-\frac{\gamma^2_0 N}{r_0}(\nu_0{}^2-\nu_K^2)\,,\nonumber \\
\frac{1}{m}f_{\rm (rad)}^{\hat \phi}&=& \frac{1}{m}[\sigma \Phi^2 {\mathcal E}(U) {\mathcal E}(n)\gamma_0^2 (\cos \beta_0 \mp \nu_0)]=0\,,
\end{eqnarray}
and are satisfied by
\beq
\pm \nu_0=\cos \beta_0=\frac{bN}{r_0} \quad \to \quad 
\gamma_0=1/|\sin \beta_0|\,,
\eeq
and
\beq
\label{moto_gen2f}
-\frac{\gamma_0^2 N}{r_0}(\nu_0{}^2-\nu_K^2) = \frac{\sigma\Phi_0^2{\mathcal E}^2}{m}\frac{\gamma_0\sin^3\beta_0}{r_0 N^2\sqrt{r_0^2-b^2N^2}} \,,
\eeq
being now
\beq
\hat \nu_k \cdot \nu=\cos^2 \beta_0 \quad \to \quad
{\mathcal E}(U)=\gamma_0 {\mathcal E}(n)\sin^2\beta_0\,.
\eeq
The equilibrium condition (\ref{moto_gen2f}) can then be rewritten as 
\begin{eqnarray}
\label{moto_gen2f3}
N \gamma_0^3 \left(1-\frac{\nu_0^2}{\nu_K^2} \right)={\rm sgn}[\sin \beta_0]\frac{A}{M}\,.
\end{eqnarray}
Clearly, when $b=0$ (i.e. $\nu_0=0$, $\gamma_0=1$) and $\sin \beta_0>0$, i.e. in the case of purely radial outward photon motion, Eq. (\ref{moto_gen2f3}) reduces to  
\begin{eqnarray}
\label{equilfin}
M N =A \qquad \to \qquad  r_0 =\frac{2M}{1-A^2/M^2}\,.
\end{eqnarray}

The stability of these orbits has been studied in detail in Ref. \cite{PR2}, to which we also refer for further analysis.

\section{Analogies between different kind of situations}

Consider the equilibrium circular orbit associated with 
different kind of particles as discussed in Section 4.

\begin{enumerate}
  \item Particles with charge $q$ in an external magnetic field 
\beq
\gamma  (\nu^{\hat \phi}{}^2-\nu_K^2)=r_0\zeta_0 \nu^{\hat \phi}\,,\qquad \zeta_0=qB_0/m\,.
\eeq
\item Particles with a magnetic dipole in an external magnetic field 
\beq
\gamma (\nu^{\hat \phi}{}^2-\nu_K^2)=-\beta\frac{M}{r_0 N }\,,\qquad \beta=2\mu B_0/m\,.
\eeq
\item Particles with spin in the background geometry  
\beq
(\nu^{\hat \phi}{}^2-\nu_K^2)=\frac{3M^2}{r_0^2 N}{\hat s}\nu^{\hat \phi}\,,\qquad  {\hat s}=s/(mM)\,.
\eeq
\item Neutral particles in a given  radiation field
\beq
N \gamma^3  \left(1-\frac{\nu_{\hat \phi}^2}{\nu_K^2} \right)={\rm sgn}[\sin \beta_0]\frac{A}{M}\,,\quad A=\sigma \Phi_0^2 {\mathcal E}^2/m \,.
\eeq
\end{enumerate}
In all these  cases (as well as in cases which are combinations of these), which originate in  different contexts, deviations from circular geodesic motion  are given by
\beq
\nu^{\hat \phi}=\pm \nu_K + \Delta \nu\,,
\eeq
where, with obvious meaning of notation
\beq
\fl\qquad
\begin{array}{lll}
\Delta \nu_{\zeta_0}&=\pm\displaystyle\frac{r_0\zeta_0}{2\gamma_K}\,,\qquad 
&\Delta \nu_{{\hat s}}=\pm\displaystyle\frac32\left(\frac{M}{r_0}\right)^{3/2}\nu_K{\hat s}\,,  \\
\Delta \nu_{\mu}&=-\displaystyle\frac{\beta}{2\gamma_K}\left(\frac{M}{r_0}\right)^{1/2}\,,\qquad
&\Delta \nu_A = \displaystyle\frac{1}{2 \gamma_K^3}\left(\frac{M}{r_0}\right)^{1/2} {\rm sgn}[\sin \beta_0]\frac{A}{M}\,.
\end{array}
\eeq
\noindent
If these  spatially circular orbits which mark the equilibrium were the object of a measurement, the uncertainty of the spatial velocity mirrors the uncertainty about the structure of the particle.
In the weak field limit, the identification of the corrections about a fixed $r_0$, implies the following kind of ambiguities, which should always be taken into account: 
\begin{enumerate}
\item  One cannot distinguish between a particle with a magnetic dipole $\mu$ moving on a mean radius $r_0$ and a one with electric charge $|q|=2\mu M^{1/2}r_0^{-3/2}$, for any mass $m$ and a magnetic (test) field $B_0$.
\par
\item  One cannot distinguish between a neutral particle moving on a mean radius $r_0$ with spin $s$ and a particle having a magnetic dipole $\mu$ in a magnetic field $B_0$ with $|\mu B_0|=(3/2)\nu_{_K}\gamma_{_K} (s/r_0)$, for any mass $m$.
\par 
\item One cannot distinguish between a spinning particle with spin $s$ on a mean radius $r_0$ and a charged particle in a magnetic field $B_0$ with $|qB_0|=3M^{1/2}r_0^{-5/2}\gamma_{_K}\nu_{_K} s$.
\par
The latter case is complementary to the previous ones. It is then clear that a measurement of the correction  to any given geodesic property, is not sufficient by itself alone to identify the structure of the particle under consideration.
Only combined measurements of different kinds can overcome this ambiguity.
\item One cannot distinguish between a spinning particle with spin $s$ also endowed with a magnetic dipole moment (e.g. a pulsar) and neutral non-spinning and not magnetized geodesic particle. The discussion about this point has been made explicitly in Section 4. 

\end{enumerate}

\section{Concluding remarks}

We have studied the geometrical (Frenet-Serret intrinsic) properties of generally non-geodesic orbits of test particles with structure, moving in the equatorial plane of the Schwarzschild spacetime. 
The analysis of the motion has been performed either numerically by a direct integration of the corresponding equations  leading to complicate patterns or by studying their analytical solutions in the special cases of equilibrium circular orbits. 
In detail, we have studied the conditions which guarantee the existence of stable spatially  circular orbits with non-Keplerian velocities maintained by particular particle's structures embedded in a black hole spacetime with added test fields.
We have explicitly considered  the cases of charged particles as well as those with a magnetic dipole in an external (test) magnetic field, of spinning particles  as well as the ones undergoing Poynting-Robertson effect due to scattering of electromagnetic radiation moving along such orbits. 
Finally we have also considered the combined effect of a intrinsic spin and of a magnetic dipole moment in an external magnetic field as expected by a pulsar-like objects.
If deviations from geodesic behaviour are the results of astrophysical measurements within a clearly specified gravitational background, then an ambiguity may arise to explain the origin of such deviations. This complication can be only overcome  by combining different kind of measurements.

\appendix

\section{Superposed electromagnetic fields to the Schwarzschild spacetime}

In this appendix we shortly reproduce the results found by Bi\v c\'ak and Janis\cite{bicak} in 1985
for the solution of a general  magnetic field superposed to the Schwarzschild background. 
In terms of a vector potential such a solution is conveniently written as
\beq
A=A_{(0)}+A_{(1)}
\eeq
where
\begin{eqnarray}
A_{(0)}&=& \frac12 B_0 r^2\sin^2\theta \rmd\phi\,,\nonumber \\
A_{(1)}&=& -B_1 \left[\sin \theta \cos \theta (r-M)(\sin \phi \rmd r+r \cos \phi \rmd \phi)\right.\nonumber \\
&& \left.+r\sin\phi[(r-2M)\cos^2\theta-M]\rmd \theta \right]\,,
\end{eqnarray}
being $B_0$ and $B_1$ arbitrary constants.
The associated electromagnetic field $F=dA$ can be represented in terms of 
 electric and magnetic fields relative to a given observer $u$; in fact, denoting such fields as
\beq
E(u)=F\rightcontract u\,, \qquad B(u)={}^*F\rightcontract u 
\eeq
where $\rightcontract$ denotes right contraction and ${}^*$ the spacetime duality operation, specifically
\beq
{}^*F_{\alpha\beta}=\frac12 \eta_{\alpha\beta\gamma\delta} F^{\gamma\delta}\,,
\eeq
we have
\beq
F=u \wedge E(u)+{}^*[u\wedge B(u)]\,.
\eeq

As measured by a static observer $u=n$ in the Schwarzschild background this electromagnetic field results a purely magnetic one, i.e. the electric field vanishes ($E(n)=0$) and 
\begin{eqnarray}
\label{Bzamogen}
B(n)&=& B_0 [-\cos\theta e_{\hat r}+N\sin \theta e_{\hat \theta}]\nonumber \\
&& -B_1 [ \sin \theta \cos \phi e_{\hat r} +N\cos \theta \cos \phi e_{\hat \theta} -N\sin \phi e_{\hat \phi}]\,.
\end{eqnarray}
In this paper we have limited our considerations to the simpler case $B_1=0$, which assures the motion to be confined to the equatorial plane.

The force on a charged test particle $U=\gamma [n+\nu]$ as in Eq. (\ref{U4vel}) and due to the above external electromagnetic field is given by $f_{\rm (em)}(U)=qE(U)$, where
\beq
E(U)_{\hat \alpha}=F_{\hat \alpha \hat \beta}U^{\hat \beta}=
\gamma [\nu \times B(n)]_{\hat \alpha}=
\gamma \eta(n)_{\hat \alpha \hat \beta \hat \gamma}\nu^{\hat \beta} B(n)^{\hat \gamma}\,;
\eeq
here $\eta(n)_{\hat \alpha \hat \beta \hat \gamma}=\eta_{\hat t \hat \alpha \hat \beta \hat \gamma}$.
In components
\begin{eqnarray}
\label{fcompsgen}
E(U)_{\hat r}&=& \gamma (\nu^{\hat \theta} B(n)^{\hat \phi}-\nu^{\hat \phi} B(n)^{\hat \theta})\,,  \nonumber \\
E(U)_{\hat \theta}&=& \gamma (\nu^{\hat \phi} B(n)^{\hat r}-\nu^{\hat r} B(n)^{\hat \phi})\,,\nonumber \\
E(U)_{\hat \phi}&=& \gamma (\nu^{\hat r} B(n)^{\hat \theta}-\nu^{\hat \theta} B(n)^{\hat r})\,.
\end{eqnarray}
The motion of a  particle (with mass $m$ and charge $q$) under the effect of this force is governed by the equations
\beq
m a(U)=f_{\rm (em)}(U)\,,
\eeq
where the components of $a(U)$ are given explicitly by Eq. (\ref{fouracc}).

\section*{Acknowledgments}

DB and AG thank Profs. R.T. Jantzen, O. Seme\v{r}\'ak and L. Stella for useful discussions on Poynting-Robertson effect around black holes.
All the authors thank ICRANET for support.

\end{document}